\documentclass[12pt,amsmath,amssymb,showkeys,prb]{revtex4}
\usepackage{setspace}
\usepackage{graphicx}

\begin{document}

$\null$
\hfill {October 30, 2018} 
\vskip 0.3in

\begin{center}
{\Large\bf Coarse-Grained Residue-Based Models of Disordered}\\ 

\vskip 0.3cm

{\Large\bf Protein Condensates: Utility and Limitations of}\\

\vskip 0.3cm

{\Large\bf Simple Charge Pattern Parameters}

\vskip .5in
{\bf Suman D{\footnotesize{\bf{AS}}}},$^{1}$
{\bf Alan A{\footnotesize{\bf{MIN}}}},$^{1}$
{\bf Yi-Hsuan L{\footnotesize{\bf{IN}}}},$^{1,2}$
 and
{\bf Hue Sun C{\footnotesize{\bf{HAN}}}}$^{1,3,*}$

$\null$

$^1$Department of Biochemistry,
University of Toronto, Toronto, Ontario M5S 1A8, Canada;\\
$^2$Molecular Medicine, Hospital for Sick Children, Toronto, 
Ontario M5G 0A4, Canada\\
$^3$Department of Molecular Genetics,
University of Toronto,\\ Toronto, Ontario M5S 1A8, Canada;\\

\vskip 1.3cm

%

\end{center}

\vskip 1.3cm

\noindent
$*$Corresponding author\\
{\phantom{$^\dagger$}}
E-mail: chan@arrhenius.med.utoronto.ca;
Tel: (416)978-2697; Fax: (416)978-8548\\
{\phantom{$^\dagger$}}
Mailing address:\\
{\phantom{$^\dagger$}}
Department of Biochemistry, University of Toronto,
Medical Sciences Building -- 5th Fl.,\\
{\phantom{$^\dagger$}}
1 King's College Circle, Toronto, Ontario M5S 1A8, Canada.\\

\noindent

\vfill\eject

\noindent
{\large\bf Abstract}\\

\noindent
Biomolecular condensates undergirded by phase separations of proteins 
and nucleic acids serve crucial biological functions.
To gain physical insights into their genetic basis,
we study how liquid-liquid phase separation (LLPS) of intrinsically 
disordered proteins (IDPs) depends on their sequence charge patterns 
using a continuum Langevin chain model wherein each amino acid residue
is represented by a single bead.  Charge patterns are 
characterized by the ``blockiness'' measure $\kappa$ and the ``sequence 
charge decoration'' (SCD) parameter.  Consistent with random phase 
approximation (RPA) theory and lattice simulations, LLPS propensity 
as characterized by critical temperature $T^*_{\rm cr}$ increases with 
increasingly negative SCD for a set of sequences showing a positive 
correlation between $\kappa$ and $-$SCD. Relative to RPA,
the simulated sequence-dependent variation in $T^*_{\rm cr}$ 
is often---though not always---smaller, whereas 
the simulated critical volume fractions are higher. 
However, for a set of sequences exhibiting an anti-correlation between 
$\kappa$ and $-$SCD, the simulated $T^*_{\rm cr}$'s 
are quite insensitive to either parameters. Additionally, we find 
that blocky sequences that allow for strong 
electrostatic repulsion can lead to coexistence curves with
upward concavity as stipulated by RPA, but the LLPS
propensity of a strictly alternating charge sequence was likely
overestimated by RPA and lattice models because interchain stabilization 
of this sequence requires spatial alignments that are difficult
to achieve in real space. These results help delineate the utility and 
limitations of the charge pattern parameters and of RPA, pointing
to further efforts necessary for rationalizing the newly observed subtleties.

\vfill\eject


\noindent
{\Large\bf 1 Introduction}\\

Functional biomolecular condensates of proteins and nucleic acids---some of 
which are referred to as membraneless organelles---have been garnering intense 
interest since the recent discoveries of liquid-like behaviors of germline 
P-granules in {\it Caenorhabditis elegans} \cite{brangwynne2009} and 
observations of phase transitions from solution to condensed liquid and/or to 
gel states in cell-free systems containing proteins with significant 
conformational disorder.\cite{Rosen12,McKnight12,Nott15,tanja2015,parker2015}
In hindsight, the possibility that certain cellular compartments were condensed 
liquid droplets has already been raised more than a century ago when the 
protoplasm of echinoderm (e.g. star-fish and sea-urchin) eggs was seen
as an emulsion with granules or microsomes as its basic 
components.\cite{wilson1899} Subsequently, in two studies nearly half a 
century apart, the nucleolus was hypothesized to be a ``coacervate'', a
``separated phase out of a saturated solution'' \cite{lars1946} and, more
generally, phase separation in the cytoplasm was proposed to be ``the basis for 
microcompartmentation''.\cite{brooks1995} Now, burgeoning 
investigative efforts on biomolecular condensates in the past few years have 
yielded many advances (refs\cite{rosen2017,cliff2017,eggert2018,dinneny2018} 
and references therein). To name a few, 
phase separations of intrinsically disordered proteins (IDPs) or 
folded protein domains connected by disordered linkers are critical 
in the formation and organization of the nucleolus \cite{Feric16} 
(as anticipated seventy years earlier \cite{lars1946}), the nuclear pore 
complex \cite{anton,anton_biochem}, postsynaptic 
densities \cite{MJZ2016,MJZbiochem}, P-granules \cite{Nott15}, and
stress granules.\cite{tanja2015,Riback_etal2017} They are also responsible
for the ability of tardigrades (``water bears'') to survive 
desiccation \cite{boothby2017} and the synthesis of squid beaks \cite{squid}
as well as byssuses for anchoring mussels onto sea rocks.\cite{mussel}
More speculatively,
the compartmentalization afforded by IDP phase separation might even 
be important in the origin of life \cite{keating1,keating2} as
envisioned in the Oparin theory \cite{oparin} and 
its modern derivatives.\cite{dysonJ,dysonB} Because of the crucial roles
of biomolecular condensates in physiological functions, their dysfunction
can lead to diseases such as pathological protein 
fibrillization \cite{tanja2015} and neurological disorders.\cite{MJZ2016}


Properties of IDPs and their phase separations are dependent, as 
physically expected, upon the amino acid sequences of 
the IDPs.\cite{Nott15,muthu-pattern,pappu13,PappuCOSB}
However, deciphering genetically encoded sequence effects on biologically 
functional biomolecular condensates is difficult in general because the
interactions within such condensates can be extremely complex, often 
involving many species of proteins and nucleic acids and the condensates
are sometimes maintained by non-equilibrium 
processes.\cite{cliff2017,Deniz17,babu2018,pappu2018}
For instance, some crucial interactions in biomolecular condensates
can be ATP-modulated as in stress 
granules \cite{ATP1}, others can be tuned by post-translational 
modifications as exemplified by phosphorylations of Fused in 
Sarcoma (FUS).\cite{fus1,fus2} Biomolecular condensates are ``active 
liquids'' in this regard.\cite{active2011,active2017,berry2018,lee2018}
Moreover, some biomolecular condensates are not entirely liquid-like but
rather exhibit gel- or solid-like characters.\cite{gelation,biochemrev}
By comparison, experimental biophysical studies often focus, for tractability, 
on equilibrium properties of simple condensates consisting of only a few 
biomolecular components. Nonetheless, although these constructs are highly 
simplified models of {\it in vivo} biomolecular condensates, knowledge 
gained from their study is extremely useful not only as a scientific 
stepping stone to understanding the workings of
complex {\it in vivo} biomolecular condensates 
but also as an engineering tool for designing bioinspired 
materials.\cite{steph17,chilkoti2015,singperry2017,chilkoti2017,rohitJMB}


Currently, theoretical approaches to sequence-dependent biophysical properties 
of biomolecular condensates are only in their initial stages of development. 
These efforts---which include analytical theories and explicit-chain 
simulations---have been focusing on general principles and rationalization of
experimental data on simple systems. Analytical theories are 
an efficient investigative tool despite their limited, approximate treatment 
of structural and energetic details.\cite{biochemrev} For example, predictions 
of mean-field Flory-Huggins (FH)-type theories \cite{Nott15,CellBiol,NatPhys}
are sensitive to IDP amino acid compositions but FH theories do not 
distinguish different IDP sequences sharing the same composition. Nevertheless, 
such theories can be very useful, as demonstrated by a recent formulation that
rationalizes how FUS phase behaviors depend on tyrosine and arginine 
compositions.\cite{moleculargrammar} By comparison, more energetic details 
of multiple-chain IDP interactions are captured by random phase 
approximation (RPA), which is an analytical formulation \cite{delaCruz2003} 
that offers a rudimentary account of sequence-dependent electrostatic effects 
on IDP phase behaviors.\cite{linPRL,linJML,lin2017,njp2017} Because it
allows for the treatment of any arbitrary sequence of charges, RPA has proven
useful in accounting
for the experimental effects of sequence charge pattern on the phase properties 
of RNA helicase Ddx4 (refs\cite{Nott15,linPRL}). It is also instrumental
in proposing a novel correlation between sequence-dependent single-chain 
properties and multiple-chain phase behaviors \cite{lin2017}---which was 
recently verified by explicit-chain simulations \cite{jeetainPNAS}---and in 
suggesting a new form of ``fuzzy'' molecular recognition based on charge 
pattern matching.\cite{njp2017} In this connection, another recent approach 
that combines transfer matrix theory and simulation has also been useful in 
accounting for complex coacervation involving polypeptides with simple 
repeating sequence charge patterns.\cite{singperry2017,lytle2017}
Building on these advances, further work will be needed to develop theories 
that can account for sequence-dependent non-electrostatic effects,
including hydrophobicity, cation-$\pi$ interactions---which play significant
roles in functional \cite{larry2014} and 
disease-causing \cite{kaw2013} IDP interactions and in the formation 
of biomolecular condensates \cite{Nott15,mussel}---as well as
aromatic \cite{diederich} and non-aromatic \cite{robert}
$\pi$--$\pi$ interactions which are likely of importance in the assembly
of biomolecular condensates.\cite{robert}

Explicit-chain models and analytical theories are complementary.
Compared to analytical theories, explicit-chain simulations of IDP phase 
separation are computationally expensive because they require tracking 
the configurations of a multiple-chain model system that is sufficiently
large to represent phase-separated states. Yet explicit-chain simulations 
are necessary for a realistic representation of chain geometry and thus 
indispensable also for evaluating the approximations invoked by analytical 
theories.\cite{suman1,suman1C} Phase separation of IDP and/or folded protein
domains connected by disordered linkers have been simulated using highly
coarse-grained models consisting of basic units each designed to represent 
groups of amino acid residues.\cite{gelation,Ruff15,harmonNJP} These 
constructs have yielded physical insights into the phase behaviors of a 
four-component system,\cite{harmonNJP} for example. Also utilized
recently for explicit-chain modeling of biomolecular condensates are
coarse-grained approaches that capture more structural and energetic 
details by representing each amino acid residue of an IDP as a single 
bead on a chain.\cite{jeetainPNAS,suman1,dignon18} Because
we are interested in biomolecular condensates in which
the protein chains are significantly disordered,\cite{fawzi2015,jacob2017}
analytical theories \cite{foffi2009,vojko2015,vojko2016} and simulation 
techniques \cite{foffi2007,foffi2011,hxzhou2018} developed 
for the phase separation of
folded proteins \cite{hxzhou2016,hxzhou2017} (e.g.
$\gamma$-crystallin \cite{crystallin} and lysozyme \cite{lysozyme,lysozymeP}) 
are not directly applicable. Our group has previously 
employed lattice models to study sequence-dependent electrostatic effects 
on IDP phase separation.\cite{suman1} To assess the extent to which 
predictions from these models are affected by lattice artifacts and to 
broaden our effort to model biomolecular condensates in general, here
we apply more realistic coarse-grained models wherein IDP chains are 
configured in the continuum.\cite{JChen2012,cosb15,Best2017,Shea2017}


Coarse-grained explicit-chain models are well-suited to
address general physical principles.
The rapidly expanding experimental efforts have provided
an increasing rich set of data on overall physical properties of
biomolecular condensates that awaits theoretical analysis.
For instance, although solutions with temperature-independent effective
solute-solute interactions are expected to phase separate when temperature 
is reduced below a certain upper critical solution temperature (UCST)---in
which case phase separation propensity at a given temperature increases with 
increasing critical temperature $T^*_{\rm cr}=$ UCST, some biomolecular 
condensates are formed at raised temperatures (i.e., they possess a lower 
critical solution temperature, LCST)---in which case phase separation 
propensity at a given temperature decreases with 
increasing $T^*_{\rm cr}=$ LCST. Examples of the latter 
include elastin,\cite{elastin1,lisa,tanja2018} the 
Alzheimer-disease-related tau protein,\cite{tau} and the Poly(A)-binding 
protein Pab1 associated with stress granules in yeast.\cite{Riback_etal2017} 
Recent experiments on elastin indicate that formation of biomolecular 
condensates can also be dependent upon hydrostatic pressure.\cite{roland18} 
As has been suggested,\cite{biochemrev,roland18}
these phenomena may be accounted for, at least semi-quantitatively, by 
temperature \cite{maria} and pressure \cite{cristiano,heinrich}-dependent
sidechain \cite{alex2018} and backbone IDP interactions.\cite{biochemrev}

Building on our recent lattice simulation,\cite{suman1} we
focus here on sequence-dependent electrostatic effects on IDP phase
separation. Previous studies by analytical theories \cite{lin2017,njp2017} 
and explicit-chain lattice simulations \cite{suman1} of IDPs with different 
charge patterns suggest
that their propensities to phase separate are well correlated with
two parameters for characterizing sequence charge pattern: 
the intuitive $\kappa$ parameter for ``blockiness'' of the charge 
arrangement along a sequence \cite{pappu13,PappuCOSB}
and the ``sequence charge decoration'' SCD parameter that arose from 
a theory for the conformational dimensions of 
polyampholytes.\cite{Sawle15,Sawle17,Firman18}
If such parameters (and even simpler properties such as the net 
charge of a sequence) can predict certain aspects of 
IDP phase separation, they may shed light on the relevant ``holistic''
physical properties underpinning certain shared
biological functions among IDP sequences that are otherwise highly diverse on 
a residue-by-residue basis.\cite{moses17,doug2017} These 
parameters could be useful for designing artificial protein polymers as
well.\cite{chilkoti2018} Remarkably, although both $\kappa$ and SCD 
originated from studies of single-chain IDP properties, they 
appear to capture also the propensities of multiple IDP chains to phase 
separate.\cite{lin2017,njp2017,suman1} In view of the prospective broad 
utility of this putative relationship, its generality 
deserves closer scrutiny.
\\


\noindent
{\Large\bf 2 Scope and rationale}\\

With the above consideration in mind, the present study compares 
polyampholytes phase properties predicted by RPA theory against those 
simulated by explicit-chain models, and assesses the ability of
$\kappa$ and SCD to capture the theoretical/simulated trends.
The interplay between the effects of charge-dependent electrostatic and 
charge-independent Lennard-Jones-type interactions on polyampholyte 
phase behaviors is also explored. 

Insofar as explicit-chain modeling of biomolecular systems is
concerned, atomic models with detailed structural and energetic
representations and coarse-grained models are complementary 
when both approaches are viable for the system in question. Despite
their relative lack of structural and energetic details---and in some
cases precisely because of this lack of details---coarse-grained models 
have contributed significantly to theoretical advances since they are 
computationally efficient tools for conceptual development and for
discovery of universality across a large class of seemingly unrelated 
phenomena. For instance, early exact enumerations of conformational 
statistics of lattice polymers \cite{domb69} was instrumental in 
the subsequent fundamental development of scaling \cite{deGennes79} and 
renormalization group \cite{freed1987} theories in polymer physics.
Other examples include lattice investigations of protein folding 
kinetics \cite{kaya03} and DNA topology \cite{Liu06} that led to 
more sophisticated models confirming insights originally gained from
earlier lattice studies.\cite{TaoPCCP,Liu15} 
Lattice models are a powerful tool for the study of homopolymer 
phase separation as well,\cite{panagio1998} although their applicability
to long heteropolymeric chains might be 
limited\cite{panagio2003,panagio2005} as has been noted.\cite{suman1}
Moving beyond the confines of lattices, here we consider model
chains configured in continuum space.

The determination of phase diagrams of IDP liquid-liquid phase 
separation (LLPS) is computationally intensive. Currently,
all-atom explicit-water molecular dynamics is not feasible for this
task. Even a recent state-of-the-art molecular dynamics study
of the liquid structure of elastin that clocked a total simulated time of
165 $\mu$s could only model a droplet of twenty seven 
35-residue elastin-like peptides and did not provide a phase 
diagram.\cite{regis17} Besides issues of computational efficiency, 
common molecular dynamics force fields are well known to be
problematic for IDPs.\cite{DavidShaw2,sarah15} Developing a
force field that is suitable for both IDPs and globular proteins
has been a major ongoing challenge.\cite{best14,sarah17,Shaw18}
 
In this context, we adapt the coarse-grained model of Dignon 
et al.,\cite{jeetainPNAS,dignon18} which in turn 
is partly based upon simulation algorithms developed for vapor-liquid 
transitions.\cite{blas2008,panag2017} This approach is promising because 
it is computationally efficient and has already provided qualitative 
and semi-quantitative account of experimental data, a notable example 
of which is a rationalization \cite{dignon18} of the experimentally 
observed variation in phase behavior among phosphomimetic mutants of 
FUS.\cite{fus2} In contrast to Monte Carlo sampling of lattice models,
this modeling setup can provide dynamic information readily. An analysis 
of mean squared displacements \cite{dignon18} has indicated that the 
condensed liquid phases in this coarse-grained model can indeed be 
liquid-like rather than solid-like aggregates.

While the goal of the present work is to lay the necessary foundation for 
extensive comparison between theory and experiment, 
our primary focus here is on comparing explicit-chain 
results against analytical theories and assessing
the effectiveness of sequence charge pattern parameters $\kappa$ and SCD
as predictors for IDP LLPS. In view of the rationalizations afforded by 
analytical theory for experiment \cite{linPRL} and the potential
utility of analytical theories and charge pattern parameters for materials 
design, it is important to ascertain the parts played by the
physical assumptions and mathematical approximations in
the success or failure of these analytical formulations. For this purpose,
we deem it best to first consider simple ``toy-model'' sequences for
the conceptual clarity they offer. One advantage of using simple
coarse-grained models is that the general principles gleaned from
our exercise may have applications beyond IDPs, including, e.g.,
protein mimetic peptoids.\cite{peptoid1,peptoid2}

As detailed in subsequent sections of this article, our investigation
indicates that although both $\kappa$ and $-$SCD correlate positively
with RPA-predicted LLPS propensities for 
polyampholytes having zero net charge but possessing different sequence charge
patterns, the corresponding correlations with LLPS
propensities simulated by coarse-grained models are less general.
These findings help delineate the utility/limitation of 
RPA as well as that of the sequence charge parameters 
$\kappa$ and SCD as LLPS predictors.
Comparisons of our results from lattice and continuum 
explicit-model simulations suggest further that the spatial
order imposed by lattice models would likely result in overestimated
LLPS propensities for IDP configured in real space.
Ramifications of these observations for ongoing development of 
theoretical and computational techniques for biomolecular
condensates are discussed below.
\\

\noindent
{\Large\bf 3 Computational details}\\

\noindent
{\bf 3.1 Continuum coarse-grained model and simulation protocol}\\

Similar to ref\cite{dignon18},
we adopt the recent algorithm in ref\cite{panag2017} for simulating 
vapor-liquid equilibrium of flexible Lennard-Jones (LJ) chains to study 
IDP LLPS. The interactions between 
LJ spheres are now identified as effective interactions (potentials 
of mean force) between amino acid residues in a liquid solvent. 
Consequently, the vapor and liquid phases
in the original formulation\cite{panag2017} correspond, respectively, to 
the dilute and condensed liquid phases of an IDP solution. 
Molecular dynamic simulations are performed with the 
HOOMD-blue \cite{anderson2008,glaser2015} simulation
package with IDP chains (polymers) configured in a cubic box
with periodic boundary conditions. The long-spatial-range electrostatic 
interaction among the charged residues (monomers) is treated by PPPM 
method implemented in the package \cite{lebard2012}.

Using the notation in our previous lattice study,\cite{suman1} for any two 
different residues labeled $\mu,i$ and
$\nu,j$ ($\mu,\nu = 1,2,\dots,n$
label the IDP chains where $n$ is the total number of chains
in the simulation, $i,j=1,2,\dots,N$ label the $N$ residues
in each chain) with charges $\sigma_{\mu i},\sigma_{\nu j}$ in units
of elementary electronic charge $e$, their
electrostatic interaction is given by
\begin{equation}
(U_{\rm el})_{\mu i,\nu j}
= \frac {\sigma_{\mu i}\sigma_{\nu j}e^2}{4\pi\epsilon_0\epsilon_{\rm r}
r_{\mu i,\nu j}}
\; , 
\end{equation}
where $\epsilon_0$ is vacuum permittivity, $\epsilon_{\rm r}$ is relative 
permittivity (dielectric constant), and
$r_{\mu i,\nu j}$ is the distance separating the two residues. 
Unlike refs\cite{suman1,dignon18}, the electrostatic interactions are
not screened in the present study. (Note that the expression for
$(U_{\rm el})_{\mu i,\nu j}$ in ref\cite{suman1} is in units of
$k_{\rm B}T$ where $k_{\rm B}$ is Boltzmann constant and $T$ is absolute 
temperature).  Besides electrostatics, all non-bonded 
residue pairs also interact via the LJ potential
\begin{equation}
(U_{\rm LJ})_{\mu i,\nu j} = 4\epsilon \left[\left(\frac {a}{r_{\mu i,\nu j}}
\right)^{12} -
\left(\frac {a}{r_{\mu i,\nu j}}\right)^6 \right] \; ,
\end{equation}
where $\epsilon$ is the LJ well depth (not to be confused with the
permittivities) and $a$ specifies the LJ interaction range. 
The electrostatic and LJ interactions in Eqs.~(1) and (2) apply
to all intra- and interchain residue pairs that are not sequential
neighbors along a chain, i.e., for all $\mu,i$ and $\nu,j$ without
exception when $\mu\ne\nu$ and for all $\mu,i$ and $\mu,j$ satisfying 
$|i-j|>1$ when $\mu=\nu$.
For simplicity and to facilitate a more direct comparison with
our previous theoretical \cite{linJML,lin2017} and lattice \cite{suman1} 
studies, we use the same $a$ for the two types of residues considered
below (unlike ref\cite{dignon18} which uses different $a$ values
for different residue types).
As suggested by previous simulations of phase 
coexistance,\cite{trokhymchuk1999,duque2004} we expect a LJ cutoff 
distance of $6a$ is adequate and thus it is adopted for our simulations.
For computational efficiency, the same cutoff is applied also to
the electrostatic interaction in Eq.~(1).
We set $\epsilon=e^2/(4\pi\epsilon_0\epsilon_{\rm r}a)$ and use $\epsilon$
to define the energy scale throughout the present study, including
cases when the LJ potential is reduced to $(U_{\rm LJ})/3$ (see below). 
All temperatures reported below are reduced 
temperature $T^*\equiv k_{\rm B}T/\epsilon$. 
(Thus $T^*$ can be converted to $T$ for any given relative 
permittivity $\epsilon_{\rm r}$, although the present theoretical analysis
largely does not focus on specific $\epsilon_{\rm r}$ values.)
The strong interactions maintaining chain connectivity are modeled by 
a harmonic potential between successive residues along a chain:
\begin{equation}
U_{\rm bond}(r_{\mu i,\mu i+1}) = K_{\rm bond} (r_{\mu i,\mu i+1}-a)^2/2 
\;
\end{equation}
where the spring constant $K_{\rm bond}=75,000\epsilon/a^2$ is similar
to corresponding values used for bond-length energies in the TraPPE force 
field.\cite{mundy1995,martin1998,nicolas2009,pamies2010}
Kinetic properties of the simulated system is modeled by Langevin dynamics
using the velocity-Verlet algorithm with a timestep of
$0.001\tau$, where $\tau \equiv\sqrt{ma^2/\epsilon}$ and $m$ 
is the mass of a residue (for simplicity all residues are assumed to have
the same mass). As in ref\cite{panag2017}, we use a weakly coupled Langevin 
thermostat with a friction factor of $0.1m/\tau$ (ref\cite{allen1991_book}). 

We begin each simulation by randomly placing $n=500$ IDP chains in 
a periodic cubic simulation box of length $70a$. Subsequently, the chain 
configurations are energy-minimized and then heated to a high $T^*=4.0$ for 
$5,000\tau$. This is followed by a compression of the periodic 
simulation box (by isotropic rescaling of all chain coordinates)
at a constant rate under the same high $T^*=4.0$ for $10,000\tau$ 
to arrive at a much smaller periodic cubic box of length
$33a$, resulting in a final IDP density 
$\rho\approx 0.7m/a^3$. The simulation box is then expanded along
the direction (labeled as $z$) of one of the three axes of the box 
by a factor of eight with the temperature kept at a low $T^*=1.0$, 
resulting in a simulation box with dimensions
$33a\times 33a\times 264a$ containing a concentration of chain 
population (a ``slab'') somewhere along the $z$-axis whereas chain 
population is zero or extremely sparse for other parts of the
elongated simulation box. Any conformation that is originally wrapped
in the $z$-direction in the compressed $33a\times 33a\times 33a$ box
because of the periodic boundary conditions is unwrapped in this
expansion process by placing the chain conformation entirely on the
side of the ``slab'' with larger $z$ values (see Fig.~1 for a 
visualization\cite{vmd} of this procedure).

After this initial preparation, the periodic boundary conditions along 
the $z$-axis are re-instated. The temperature of the expanded simulation 
box is changed from $T^*=1.0$ to the temperature of interest and equilibrated
for $30,000\tau$. The production run is then carried out for 
$100,000\tau$ during which snapshots of the chain configurations are
saved every $10\tau$ for detailed analyses. The position of the simulation
box is continuously adjusted such that the center of mass of the chains
is always at $z=0$. Density distributions
along the $z$ axis are determined by averaging subpopulations of 264 bins of
equal width ($=a$) over the simulated trajectories.\cite{panag2017} 
Polyampholytes densities are reported in units of $m/a^3$. It follows
that the numerical value of $\rho$ is equal to the average
number of residues (monomers) in a volume of $a^3$.
An example of the results from such a calculation is given in Fig.~1.
\\

\noindent
{\bf 3.2 Sequence charge pattern parameters}\\

Following Das and Pappu,\cite{pappu13} the blockiness parameter
$\kappa$ is defined to quantify the deviations of the charge
asymmetries of local sequence segments from the overall charge asymmetry 
of a given sequence. For a sequence segment of length $g$ that
starts at monomer $k$ (on any one of the $n$ identical chains
labeled by $\mu$), the charge asymmetry is defined as 
$s(g;k) = [f_+(g;k) - f_-(g;k)]^2/[f_+(g;k) + f_-(g;k)]$
where $f_+(g;k)$ and $f_-(g;k)$ are the ratios, respectively, of positively and 
negatively charged monomers (residues) among the $g$ monomers of the
sequence segment; i.e., 
$f_\pm = \sum_{i=k}^{k+g-1} (|\sigma_{\mu i}| \pm \sigma_{\mu i})/2g$
where the summation is over the sequence segment that starts at
monomer $k$ and ends at monomer $k+g-1$.
It follows that the overall charge asymmetry for the entire
sequence with $N$ monomers is $s(N;1)$.
The average deviation of local charge asymmetry
from the overall charge asymmetry for all $g$-monomer segments
(sliding windows) is given by 
$\delta_g\equiv\sum_{k=1}^{N-g+1}[s(g;k)-s(N;1)]^2/(N-g+1)$.
A $g$-specific quantity $\kappa_g \in [0,1]$ is then defined as
$\kappa_g\equiv \delta_g/\max(\delta_g)$ where $\max(\delta_g)$
is the maximum $\delta_g$ value of the set of sequences with a given
composition that is being considered.\cite{pappu13} In the present case,
$\max(\delta_g)$ corresponds to the $\delta_g$ of the fully charged 
$N$-monomer diblock polyampholyte. As in ref\cite{pappu13}, the $\kappa$
we have used for the present work, which takes the form
\begin{equation}
\kappa \equiv \frac{\delta_5+\delta_6}{{\rm max}(\delta_5)+{\rm max}(\delta_6)}
\; ,
\end{equation}
is an average over results for local segment lengths $g=5$ and $g=6$.
Note that Eq.~(4) differs slightly from the $\kappa=(\kappa_5+\kappa_6)/2$ 
definition in ref\cite{pappu13} but the difference 
is practically negligible ($<1\%$ for low-$\kappa$ sequences and 
$<0.01\%$ for large-$\kappa$ sequences).

Following Sawle and Ghosh,\cite{Sawle15}
\begin{equation}
{\rm SCD} \equiv \sum_{i=2}^N \sum_{j=1}^{i-1} \sigma_{\mu i}\sigma_{\mu j}
\sqrt{i-j}/N
\end{equation}
is the weighted summation over all pairs of charges along a given sequence.
\\

\noindent
{\bf 3.3 Selection of model sequences}\\

We study seven fully charged polyampholyte sequences of length $N=50$.
The sequences have equal number of positive and negative residues 
(charge $\sigma=\pm 1$). Following the nomenclature used in previous 
studies,\cite{pappu13,lin2017,njp2017,suman1,Sawle15} 
we designate the positive and negative residues as
``lysine'' (K) and ``glutamic acid'' (E), respectively. The sequences
are referred to as ``KE'' sequences (Fig.~2). Sequences labeled
as sv1, sv15, and sv30 were originally introduced in ref\cite{pappu13}
and have been studied previously by theory\cite{lin2017,njp2017,Sawle15}
and explicit-chain simulations.\cite{pappu13,jeetainPNAS,suman1}
These sequences are chosen again for the present study because they span
a wide range of values for the sequence charge pattern parameters 
$\kappa$ and SCD. To provide a context for our simulation study, we
have examined the distributions of SCD and $\kappa$ among all
possible KE sequences with zero net charge by using simple Monte
Carlo as well as Wang-Landau \cite{WL1,WL2} sampling. The results in Fig.~3a,b
indicate that the distributions are concentrated in relatively small 
$\kappa$ and $-$SCD values. Sequences with large $\kappa$ or large $-$SCD
values are extremely rare. A reasonable positive correlation exists between
$\kappa$ and $-$SCD; but there is also considerable scatter
(Fig.~3c, blue circles), underscoring that the two parameters
address similar as well as significantly different sequence properties.
Fig.~3c indicates that sv1, sv15, and sv30 lie in a region where
$\kappa$ and $-$SCD are well correlated.

RPA theory\cite{lin2017} for sv1, sv15, sv30 and 27 other sv 
sequences\cite{pappu13} stipulates that LLPS propensity is well
correlated with $\kappa$ and $-$SCD. This prediction is supported to a
limited degree by explicit-chain simulation.\cite{suman1}
In view of these findings, it would be instructive
to probe the effectiveness of these charge pattern parameters as
LLPS predictors by extending our analysis to outlier sequences that do
not exhibit a positive correlation between $\kappa$ and $-$SCD.
Because such sequences likely reside in sparely populated regions
of sequence space, we use a biased sampling procedure
to locate them by maximizing the scoring function 
\begin{equation}
E\equiv 
A\left [ -{\rm SCD}/(-{\rm SCD})_{\rm max} - \kappa \right]^2 + 
h_{\rm SCD}\left[-{\rm SCD}/(-{\rm SCD})_{\rm max}\right] 
+ h_{\kappa}\kappa
\; 
\end{equation}
for KE sequences, where $A$, $h_{\rm SCD}$, and $h_{\kappa}$ are 
tunable parameters. When $E$ is maximized, the first term in
Eq.~(6) maximizes the difference
between a rescaled $-$SCD and $\kappa$ 
($-{\rm SCD}/(-{\rm SCD})_{\rm max}$,
$\kappa$ $\in [0,1]$), whereas the second and third terms control
whether a high $-$SCD or a high $\kappa$ value is preferred. 
Starting with an initial KE sequence, an exchange between a randomly 
chosen pair of K and E is attempted at each Monte Carlo step. 
The attempted exchange is accepted if it results in an increase in $E$.
Otherwise it is rejected.
Partially optimized sequences are generated in this manner
by 1,000 Monte Carlo steps.
By tuning the $A$, $h_{\rm SCD}$, and $h_{\kappa}$ parameters, we have
generated four sequences---labeled by as1, as2, as3, and as4 (Fig.~2)---that 
collectively exhibit an anti-correlation trend between $\kappa$
and $-$SCD (Fig.~3c, orange circles). The $\kappa$ and SCD values for
these sequences and those for the sv1, sv15, and sv30 sequences are summarized 
in Table 1. 
\\

\noindent
{\Large\bf 4 Results and Discussion}\\

\noindent
{\bf 4.1 Background residue-residue attraction enhances overall LLPS propensity 
but attenuates the sensitivity of LLPS to charge pattern variation}\\

For reasons to be expounded below, we consider three different 
combinations of the electrostatic [$U_{\rm el}$ in Eq.~(1)] and 
LJ [$U_{\rm LJ}$ in Eq.~(2)] potentials as the total residue-residue 
interaction energy $U$ (Fig.~4): (i) simple sum of the two terms, viz., 
$U=U_{\rm el}+U_{\rm LJ}$ (Fig.~4a); 
(ii) sum of the electrostatics term and a LJ term reduced to 1/3 of
its strength, viz., $U=U_{\rm el}+(1/3)U_{\rm LJ}$ (Fig.~4b); and
(iii) sum of the electrostatics term and a LJ term that applies only
to $r\le a$, where $r$ is the residue-residue distance, viz.,
$U=U_{\rm el}+U_{\rm LJ}$ for $r\le a$ and $U=U_{\rm el}$ for $r>a$.
We are interested in various combinations of $U_{\rm el}$ and $U_{\rm LJ}$
because they bear on one of the formulations used in a 
general explicit-chain simulation approach to study LLPS of 
IDPs.\cite{dignon18} Here, the ``with LJ'' model (Fig.~4a)
represents a somewhat extreme case in which the LJ attraction is
sufficiently strong such that the total interaction remains 
attractive when the two like charges are in close proximity 
(for $r\approx 2^{1/6}a$). 
To address the role of the background LJ interactions on 
LLPS, the ``with 1/3 LJ'' model (Fig.~4b) reduces
LJ attraction but the overall repulsion between like charges
is still considerably weaker than the attraction between opposite charges.
In contrast, the ``with hard-core repulsion'' model (Fig.~4c) retains
only the repulsive part of the LJ potential up to the residue-residue
separation at which $U_{\rm LJ}=0$ (when $r=a$), such that the strength
of repulsion between like charges is equal to that of attraction between 
opposite charges at $r=a$. This model represents an extreme case
in which attractive van der Waals interactions play no role in LLPS.  
Notably, the symmetry between repulsive and attractive 
interaction strengths and the treatment of hard-core excluded volume afforded 
by this model resemble those in RPA theory\cite{linPRL,linJML} (at least 
conceptually) and in explicit-chain lattice 
simulations.\cite{suman1} It follows that the model potential in Fig.~4c
is useful for assessing RPA and lattice results.

The phase diagrams for sequences sv1, sv15, and sv30 are calculated
using both the ``with LJ'' and ``with 1/3 LJ'' models (Fig.~5). All
simulated data points 
in the phase diagrams in this figure and subsequent figures are obtained 
directly from the density distributions of expanded simulation boxes except 
the critical points (at the top of each of the coexistence curves) are 
estimated using the scaling relation specified by Silmore et al.\cite{panag2017}
Representative chain configurations above and below the critical temperature 
are provided by the snapshots in Fig.~5.  As expected, the model chains 
exist in a single phase above the critical temperature with essentially
uniform polyampholyte density throughout the simulation box 
(Fig.~5, {\it bottom left}). In contrast, a condensed phase
(well-defined localized slab in the simulation box) persists
below the critical temperature (Fig.~5, {\it bottom right}). 
Consistent with RPA theory \cite{njp2017} and lattice simulations,\cite{suman1}
the critical temperatures [$T^*_{\rm cr}({\rm sv1})$, 
$T^*_{\rm cr}({\rm sv15})$, 
and $T^*_{\rm cr}({\rm sv30})$] of the three sequences
exhibit a clear increasing trend with increasing $\kappa$
($=0.0009$, $0.1354$, and $1.000$, respectively)
as well as increasing $-$SCD 
($=0.413$, $4.349$, and $27.84$, respectively, see Table~1)
for both the ``with LJ'' (Fig.~5a) and ``with 1/3 LJ'' (Fig.~5b) models.
More specifically, $T^*_{\rm cr}({\rm sv1})$, $T^*_{\rm cr}({\rm sv15})$,
and $T^*_{\rm cr}({\rm sv30})$ equals, respectively,
$3.52$, $3.86$, and $4.97$ in Fig.~5a and
$1.20$, $1.52$, and $3.44$ in Fig.~5b.

We expect LLPS propensities to be generally higher in the ``with LJ'' 
model (Fig.~5a) than in the ``with 1/3 LJ'' model (Fig.~5b) 
because the former model provides a stronger overall residue-residue 
attraction. This expectation is confirmed by the results in Fig.~5
showing that the $T^*_{\rm cr}$'s in Fig.~5a are substantially higher
than the $T^*_{\rm cr}$'s for the corresponding sequences in Fig.~5b.
However, the differences in LLPS properties among the three
sequences are more pronounced in the ``with 1/3 LJ'' model than in
the ``with LJ'' model. Whereas the difference 
$T^*_{\rm cr}({\rm sv15})-T^*_{\rm cr}({\rm sv1})$
in the ``with LJ'' model ($=0.34$) is nearly equal to that
in the ``with 1/3 LJ'' model ($=0.32$), the difference
$T^*_{\rm cr}({\rm sv30})-T^*_{\rm cr}({\rm sv15})$
is substantial smaller in the ``with LJ'' model ($=1.11$)
than in the ``with 1/3 LJ'' model ($=1.92$). This trend is even more 
clear when the ratios of $T^*_{\rm cr}$'s of different sequences are
compared: $T^*_{\rm cr}({\rm sv15})/T^*_{\rm cr}({\rm sv1})=1.097$
for the ``with LJ'' model, which is smaller than
the corresponding ratio of $1.267$ for the ``with 1/3 LJ'' model; and
$T^*_{\rm cr}({\rm sv30})/T^*_{\rm cr}({\rm sv15})=1.288$
for the ``with LJ'' model, which is substantially smaller than
the corresponding ratio of $2.263$ for the ``with 1/3 LJ'' model.
These results illustrate that variations in LLPS propensity
induced by different sequence charge patterns can be partially
suppressed by background residue-residue attraction
that pushes the chain molecules to behave more like homopolymers.

Interestingly, the coexistence curve for sv30 in Fig.~5b exhibits
clearly an inflection point on the condensed (right-hand) side
such that part of the coexistence curve on this side is concave upward.
A hint of upward concavity exists also---though barely discernible--for
the coexistence curve for sv30 in Fig.~5a as well as the coexistence
curves for sv15 and sv1 in 
Fig.~5b. In contrast, the entire coexistence curves for sv15 and sv1 in
Fig.~5a is convex upward. This observation from explicit-chain simulations
are consistent with RPA theory of polyampholytes with zero or near-zero 
net charge.\cite{linPRL,linJML,lin2017} Indeed, a systematic RPA study
of 30 KE sequences indicates that upward concavity of the condensed side
of the coexistence curve decreases with decreasing $-$SCD and 
decreasing $\kappa$ (Figure~1a of ref\cite{lin2017}). Whereas the
RPA-predicted concavity is prominent for sv30, it is barely discernible
for sv15 and sv1 (Figure~10 of ref\cite{suman1}). This upward concavity
of coexistence curves is known to be related to the long spatial
range of electrostatic interactions and has been predicted by RPA
theory for polyelectrolytes.\cite{delaCruz2003} Apparently---and not
inconsistent with intuition, LLPS properties of polyampholytes with 
more blocky sequence charge patterns are in some respect akin to 
those of polyelectrolytes. Comparison of the coexistence curves
in Fig.~5a against those in Fig.~5b suggests further that upward
concavity of the coexistence curve is likely associated with the presence 
of strong long-range repulsive interactions in the system as well.
In this regard, it is instructive to note that none of the
coexistence curves simulated recently in refs\cite{jeetainPNAS,dignon18}
for various intrinsically disordered proteins or protein regions
exhibit upward concavity. The only coexistence curve in these references
that shows a clear upward-concave trend is the one for a model
folded helicase domain in Figure~S14 of ref\cite{dignon18}.
\\

\noindent
{\bf 4.2 Sequence charge pattern parameters $\kappa$ and SCD are good 
predictors of LLPS propensity for some but not all polyampholytes}\\

As a group, the as1--4 sequences 
exhibits anti-correlation between $\kappa$ and $-$SCD. In contrast
to sequences sv1, sv15, and sv30 in Fig.~5 with $T^*_{\rm cr}$ 
increasing with both increasing $\kappa$ and increasing $-$SCD, the
phase diagrams for sequences as1, as2, as3, and as4 in Fig.~6 are
quite similar despite their very diverse $\kappa$ values ranging
from $0.1761$ for as1 to $0.7783$ for as4 (Table~1). Their $T^*_{\rm cr}$'s 
are 2.25, 2.31, 2.28, and 2.41, respectively. Although $T^*_{\rm cr}$
generally increases with $\kappa$ (except for as2 and as3), the
increase of $T^*_{\rm cr}$ with respect to $\kappa$ is small: From
as1 to as4, only a difference of 
$T^*_{\rm cr}({\rm as4})-T^*_{\rm cr}({\rm as1})=0.16$
and a ratio of 
$T^*_{\rm cr}({\rm as4})/T^*_{\rm cr}({\rm as1})=1.071$
are registered for an increase in $\kappa$ of $0.6022$.
By comparison, even though the difference in $\kappa$ is much smaller
at $0.1375$ for the sv1 and sv15 sequences, their $T^*_{\rm cr}$ difference 
and ratio simulated using the same ``with 1/3 LJ'' model (Fig.~5b), 
$0.32$ and $1.267$ respectively, are much larger than those between 
as1 and as4 in Fig.~6.

Because of the anti-correlation between $\kappa$ and $-$SCD among
sequences as1--4 (Fig.~3c), the $T^*_{\rm cr}$'s of 
the as1--4 sequences in Fig.~6 anti-correlate with their $-$SCD 
values---rather than correlating with $-$SCD as in the case of 
the sv1, sv15, and sv30 sequences. Specifically,
the increase of the critical temperature from as1 to as4, 
$T^*_{\rm cr}({\rm as4})-T^*_{\rm cr}({\rm as1})=0.16$,
is accompanied by a decrease in the value of $-$SCD 
from $12.79$ for as1 to $6.11$ for as4 (a difference of $6.68$).
This magnitude of the rate of change of $T^*_{\rm cr}$ with
respect to SCD is only about a third of that between sequences sv1 and sv15
and is in the opposite direction ($0.32$ change in $T^*_{\rm cr}$ from
sv1 to sv15 is concomitant with a $-$SCD increase of $3.936$).

The comparison between the results in Fig.~5 and Fig.~6 thus indicates
that $\kappa$ and SCD are sensitive predictors of the LLPS of a certain
class of polyampholytes (such as the sv1, sv15, and sv30 sequences) but 
not others (such as the as1, as2, as3, and as4 sequences). This limitation
of the $\kappa$ and SCD parameters is not entirely surprising in
view of their origins as intuitive \cite{pappu13} and theoretial \cite{Sawle15} 
predictors of single-chain conformational dimensions of polyampholytes,
not as predictors for LLPS. By construction, $\kappa$ quantifies
the degree to which the sequence charge distribution is locally blocky,
whereas SCD addresses complementarily sequence-nonlocal effects from
charges that are separated by a long segment
of the chain. For the original set of 30 polyampholytes
introduced in ref\cite{pappu13} (which includes sv1, sv15, and sv30),
SCD correlates better with explicit-chain simulated radius of 
gyration \cite{pappu13} and RPA-predicted $T^*_{\rm cr}$'s.\cite{lin2017}
Now, the $T^*_{\rm cr}$'s weak positive correlation with $\kappa$
and weak negative correlation with $-$SCD for the as1--4 sequences in Fig.~6 
suggest that the effect of local charge pattern on LLPS---which is a 
multiple-chain phenomenon---may be stronger than that of nonlocal charge 
pattern. Nonetheless, the fact that $\kappa$ as a LLPS predictor
is much less sensitive when it anti-correlates with $-$SCD suggests
at the same time that nonlocal charge pattern effect does have a non-negligible
role in LLPS.  We will return to this issue below when
we present an extensive study of these sequence charge parameters 
in the context of RPA theory in Sec. 4.5.
\\

\noindent
{\bf 4.3 LLPS of polyampholytes in the absence of background non-electrostatic
residue-residue attraction may require highly segregated charge patterns}\\

To examine further the effect of background LJ interactions on
the sensitivity of LLPS to sequence charge pattern, 
simulations of sv1, sv15, and sv30 are conducted using the ``with 
hard-core repulsion'' model potential in Fig.~4c. Results from this
model (Fig.~7) should be more directly comparable with those from pure RPA 
theory \cite{njp2017} (without any Flory $\chi$ parameter \cite{linJML}) 
and lattice simulations \cite{suman1} because there is
no non-electrostatic attraction in pure RPA theory and our 
recent explicit-chain lattice model.\cite{suman1} Aside from chain
connectivity and lattice constraints, the only non-electrostatic 
interactions in those formulations \cite{njp2017,suman1}
are excluded-volume repulsions.
Despite the sharp repulsive forces entailed by this model potential,
no erratic dynamics was observed in our Langevin simulations. Nonetheless,
it would be instructive in future investigations to assess more broadly 
the effects of strong intra- and interchain repulsion on phase properties 
by using Monte Carlo sampling.

Figures~7a,b show the equilibrium density distributions simulated at 
an extremely low temperature of $T^*=0.001$ for sequences sv1 and sv15. 
This temperature is approaching the lowest that can be practically 
simulated in the current model, because it is close to the minimum 
temperature fluctuation that can be maintained by the model thermostat. 
For example, we have attempted to set the thermostat to $T^*=0.0001$ but 
the actual temperature returned by the simulation was $T^*=0.002$.
Although the density distributions for sv1 and sv15 in Fig.~7a,b are not 
uniform throughout their respective simulation boxes---indicating
that the chains are to a degree favorably associated with one
another, the distributions in Fig.~7a,b do not indicate a clear signature 
of phase separation,\cite{dignon18,panag2017} namely a localized, 
well-defined slab of essentially uniform density (Fig.~1, {\it right}).
Because a temperature as low as $T^*=0.001$ is very unlikely to be
physically realizable for a liquid aqueous solution ($T^*=0.001$ 
corresponds\cite{njp2017} to $T\approx 0.5$ K for $\epsilon_{\rm r}=80$
and $T\approx 44$ K for $\epsilon_{\rm r}=1$), 
we may conclude from Fig.~7a,b that for practical purposes sv1 
and sv15 do not undergo LLPS in aqueous solutions in the absence of
substantial non-electrostatic attractive interactions.

In contrast, a clear signature of phase separation is indicated for 
sequence sv30 at a sufficiently low temperature of $T^*=0.1$ (Fig.~7c). 
The simulated phase diagram of sv30 is shown in Fig.~7d. Because of
the reduced inter-residue (thus inter-chain) attraction of the 
``with hard-core repulsion''
model (Fig.~4c) relative to the other two model potentials in Fig.~4a,b, the 
critical temperature $T^*_{\rm cr}=1.65$ for sv30 here is lower than
the $T^*_{\rm cr}$ values of $4.97$ and $3.44$ for sv30
in Fig.~5a and Fig.~5b. The upward concavity of the condensed side of 
the coexistence curve for sv30 is remarkably more prominent in Fig.~7d than 
in Fig.~5a and Fig.~5b, buttressing our contention above that this
hallmark feature is closely related to the presence of strong 
repulsive electrostatic interactions in the system.

The dramatic differences in LLPS propensity among 
the three systems studied in Fig.~7 are illustrated by two extreme 
cases of a particular energetically favorable configuration for a pair 
of sv1 chains (Fig.~8a) and one for a pair of sv30 chains (Fig.~8b).
In these configurations, inter-chain distances between contacting beads
are constant at $r=a$ and thus repulsive LJ energies do not contribute
to the total interaction energies plotted in Fig.~8c, which is given by
$\sum_{i,j; r_{\mu i,\nu j}\le r_{\rm max}} (U_{\rm el})_{\mu i,\nu j}$,
where $\mu,\nu$ are the labels for the two chains in each pair.
Figure~8c shows that the sv30 pair is energetically much more
favorable than the sv1 pair. At the large cutoff limit 
($r_{\rm max}\to \infty$), the interaction energy for the sv1 pair
limits to $-6.5\epsilon$, whereas that for
the sv30 pair limits to $-228.2\epsilon$.
This difference helps rationalize the lack of LLPS for sv1 and the
possibility of LLPS for sv30. The strictly alternating charge pattern of 
sv1 leads to a very weak net favorable interaction between a 
sv1 pair even when the pair is in the highly special---and thus 
unlikely---configuration in Fig.~8a. This is because of 
numerous partial cancellations of attractions between a pair of opposite 
charges and repulsions between a pair of like charges since such pairs 
are positioned next to each other. 
The weakness of the net favorable inter-chain interaction means that
the inter-chain attraction in a highly constrained configuration that
can readily be overwhelmed by increased configurational entropy
in an ensemble of more open chains. By comparison,
for sv30, because of the much stronger net favorable inter-chain
interaction, a condensed phase can ensue at a sufficiently low 
temperauture when the free-energy effect of configurational
entropy is relatively diminished.
\\

\noindent
{\bf 4.4 Lattice models can overestimate LLPS propensity because of
their artefactual spatial order}
\\

To gain further insight into low-temperature LLPS properties of sv1, a 
snapshot of sv1 configurations simulated at $T^*=0.001$ is shown in Fig.~9.
The chains are loosely associated but they do not coalesce into a 
droplet or a slab in the simulation box. Even the more densely populated
region of the simulation box contains region of substantial pure
solvent volumes (solvent-filled cavities or ``voids'' in the model) with 
no sv1 chains, indicating that the associated state has a very weak effective 
surface tension and is not liquid-like. The snapshot shows that
some individual chain conformations are elongated, presumably to achieve 
more favorable inter-chain contacts by near parallel alignment (similar to
Fig.~8a), but others appear more globular (Fig.~9, {\it bottom}). 
The geometric/configurational difference between the type of associated
states in Fig.~9 and unambiguously phase-separated condensed phases such 
as the one depicted in Fig.~5 ({\it bottom right}) may be quantified by the 
analysis of cavity distributions in Fig.~10, which shows 
by two different rudimentary measures of cavity size that there are
substantially more large solvent-filled cavities in the peculiar
associated state in Fig.~9 than in a condensed phase that has clearly 
undergone phase separation.

In contrast to the present continuum simulation results, 
both sv1 and sv15 were observed to coalesce into a
condensed phase in our previous explicit-chain lattice simulation \cite{suman1}
(Fig.~11). Thus, by comparing explicit-chain simulation results from the 
present continuum model against those from our previous lattice model, 
it is clear that the spatial order
imposed by the lattice can have a very significant effect in favoring
phase separation in lattice model systems. Lattice constraints
represent a significant restriction on configurational freedom, allowing
opposite charges along polyampholytes to align more optimally. This
effect is illustrated by the snaphots for condensed phases in Fig.~11.
The above observation implies that lattice models of 
phase separation can drastically overstimate phase separation 
propensity in real space. However, in some applications, it
may be argued that the lattice order can serve to mimic
certain physically realistic local configurational order---such as that 
induced by hydrogen bonding in protein secondary structure---that is 
not taken into account in a coarse-grained continuum chain 
model.\cite{chandill90,cohen1994,yee1994} Chains configured on
lattices may also capture certain effects of steric constraints such 
as those embodied in the tube model of proteins \cite{banavar2000}
(see footnote 2 on p.~S309 of ref~\cite{wallin2006}). The degree to which
these subtle ramifications of lattice features can be exploited 
in the study of IDP LLPS remains to be explored. Taken together,
these considerations indicate that lattice models can be useful
in exploring general principles (Sec. 2) and deserve further
attention in future studies; but their predictions
should always be interpreted with extra caution.
\\

\noindent
{\bf 4.5 RPA theory is useful for physically rationalizing
polyampholyte LLPS but has its limitations}
\\

We utilize the simulated  phase properties of the several polyampholyte 
sequences computed using different model potentials 
to assess predictions offered by RPA theory. To set the stage,
we first establish a broader context of RPA predictions than
is currently available. Applying the salt-free RPA 
formulation for IDP LLPS \cite{linPRL} that we adapted \cite{delaCruz2003} 
and detailed \cite{lin2017} recently, we numerically calculate
the critical temperature $T^*_{\rm cr}$ and critical volume fraction
$\phi_{\rm cr}$ of all 10,000 randomly sampled sequences in Fig.~3 and
examine their relationship with the sequence charge parameters
$\kappa$ and $-$SCD (Fig.~12). Consistent with previous observations
based on more limited datasets,\cite{njp2017,suman1} RPA-predicted
$T^*_{\rm cr}$ of polyampholytes with zero net charge exhibits a very good
correlation with $-$SCD (tight scatter in Fig.~12a) but a lesser
though still substantial correlation with $\kappa$ (broader scatter in 
Fig.~12b). The RPA-predicted spread of the $\phi_{\rm cr}$ versus $-$SCD scatter
for the same set of polyampholytes (Fig.~12c) is also narrower
than the corresponding spread of the $\phi_{\rm cr}$ versus $\kappa$ scatter
(Fig.~12d); but this difference in scatter between $-$SCD 
and $\kappa$ is not as pronounced as the corresponding difference
in the scatter for $T^*_{\rm cr}$ (Fig.~12a,b).

The RPA-predicted dependence of $T^*_{\rm cr}$'s and $\phi_{\rm cr}$'s of
the sv1, sv15, sv30 sequences on $-$SCD and $\kappa$ (red squares in Fig.~12) 
is well within the general, most probable trend expected from the 
10,000 randomly sampled sequences (blue circles Fig.~12). 
However, the as1, as2, 
as3, and as4 sequences (orange circles in Fig.~12) appear to be outliers.
These sequences' deviation from the most probable trend is mild for the
$T^*_{\rm cr}$ versus $-$SCD (Fig.~12a), 
the $\phi_{\rm cr}$ versus $-$SCD (Fig.~12c),
and the $\phi_{\rm cr}$ versus $\kappa$ (Fig.~12d) scatter plots, but is severe
for the $T^*_{\rm cr}$ versus $\kappa$ scatter plot (Fig.~12b). It is
clear from Fig.~12b that the sign of correlation
of the $T^*_{\rm cr}$'s of the as1, as2, as3, and as4 sequences
is opposite to the overall trend for the 10,000 randomly sampled
sequences (Fig.~12b).

To compare RPA predictions with explicit-chain simulation results,
we first summarize the simulation data, by themselves, in Fig.~13,
which is the simulation equivalent of the theoretical data in Fig.~12a,b.
It provides the dependence of simulated $T^*_{\rm cr}$ on
the two sequence charge pattern parameters. 
Figure~13 recapitulates the positive correlation of the simulated 
$T^*_{\rm cr}$'s of the sv1, sv15, and sv30 sequences with $-$SCD 
and $\kappa$ (squares in Fig.~13). The trends for $-$SCD and $\kappa$ 
are quite similar. However, in relative terms, the $T^*_{\rm cr}$'s of the 
as1, as2, as3, as4 sequences are almost independent of either $-$SCD 
and $\kappa$ (circles in Fig.~13). As noted above, the correlation of 
the $T^*_{\rm cr}$ of these four sequences as a set is slightly 
negative with $-$SCD and only slightly positive with $\kappa$.
Only one data point is available in each panel of Fig.~13
for simulated $T^*_{\rm cr}$ in the ``with hard-core repulsion'' model 
(diamond for sv30) because sv1 and sv15 fail to phase separate 
unequivocally in this model (see above).
The $T^*_{\rm cr}$ of sv30 in this model is similar to that of the 
less-blocky sv15 sequence in the lattice model (red-filled black squares).
As stated previously, no simulated $T^*_{\rm cr}$ is available for sv30 in our
recent lattice model because the favorable interactions in sv30 were too
strong for efficient equilibration in that model.\cite{suman1}

We now contrast our simulation data with theoretical predictions. Depending
on the simulation conditions, different matching theoretical formulations
are used for the comparison:
(i) Pure RPA theory for electrostatic and excluded-volume interactions only,
as described in ref\cite{linPRL}, is utilized to compare with present
simulations using the ``with hard-core repulsion'' potential that
does not include any non-electrostatic attraction (Fig.~4c).
(ii) The RPA+FH theory prescribed by Equation~10 in ref\cite{linPRL} with 
a Flory parameter $\chi=(2\sqrt{2}\pi/3)/T^*$ is adopted to compare with
simulations using the ``with LJ'' potential (Fig.~4a). Here the $\chi$
parameter is purely enthalpic. It is introduced to mimic the 
background enthalpic LJ interaction in the simulations, viz.,
$\chi=({\rm pairwise\ LJ\ energy})\times ({\rm pairwise\ contact\ volume})/(2k_{\rm B}T)$.
We approximate pairwise LJ energy by the well depth $\epsilon$, and the 
pairwise contact volume by that of a sphere with radius $2^{1/6}a$ which
is the residue-residue separation at which the LJ energy is $\epsilon$.
These approximations lead to $\chi=2 \epsilon\sqrt{2}\pi/(3k_{\rm B}T)$
$=(2\sqrt{2}\pi/3)/T^*$ because $T^*= k_{\rm B}T/\epsilon$
and the volume of the conceptual lattice unit for the Flory-Huggins 
consideration is $a^3$.
(iii) The same RPA+FH theory but with $\chi=(2\sqrt{2}\pi/9)/T^*$, i.e.,
1/3 of the background interaction strength, is applied accordingly to 
compare with simulations using the ``with 1/3 LJ'' potential (Fig.~4b).
(iv) RPA theory for a  screened Coulomb potential, as specified
by Equations~2 and 3 of ref\cite{suman1}, is used to compare against lattice 
simulation results for sv1 and sv15 we computed previously using screened 
electrostatics.\cite{suman1}

Predictions by these theoretical formulations are summarized
in Table~2 together with their corresponding simulation results. 
The theoretical and simulated critical temperatures and critical 
volume fractions are plotted in Fig.~14. In this
theory-simulation comparison, we stipulate that the 
polyampholyte volume fraction $\phi$ in the simulations may be identified, 
roughly, to the simulated residue density $\rho$ in Table~2 
(hence $\phi_{\rm cr}\approx \rho_{\rm cr}$). 
By definition, $\rho$ is the average 
number of residues in a volume of $a^3$, thus $\phi\propto\rho$, 
and the volume of a residue (a bead in the polyampholyte 
chain model) is $\approx 0.74a^3$ (volume of a sphere of radius $2^{1/6}a/2$).
It follows that when there is one residue per $a^3$ on average (i.e.,
when $\rho=1$), approximately $0.74$ of the system volume is occupied by 
van der Waals spheres.
With this in mind, since the maximum achievable packing
fraction of equal-sized spheres is $\pi/\sqrt{18}\approx 0.74$ also, the
maximum packing possible in our simulation system is characterized 
by $\rho\approx 1$. 
Thus, $\rho$ is already given in a unit such that it corresponds 
approximately to the  
volume fraction $\phi$ ($0\le\phi\le 1$) in Flory-Huggins theory.

Figure~14 shows that the simulated $T^*_{\rm cr}$ and $\phi_{\rm cr}$
are reasonably correlated with their theoretical counterparts for
the sv1, sv15, and sv30 sequences under various simulation conditions.
The scatter plots in Fig.~14 suggest two rough scaling relations
between simulated (``sim'') and theoretical (``thr'') quantities: 
$T^*_{\rm cr,sim}\sim$ $(T^*_{\rm cr,thr})^{0.39}$ and
$\phi_{\rm cr,sim}\sim$ $(\phi_{\rm cr,thr})^{0.19}$.
The data points for different models in Fig.~14 indicate clearly 
that these relations hold quite well for the simulated $T^*_{\rm cr}$'s 
and simulated $\phi_{\rm cr}$'s of sv1, sv15, and sv30 
for the ``with LJ'' and ``with 1/3 LJ'' potentials but they fit 
poorly with the simulation data of as1, as2, 
as3, and as4 (for the ``with 1/3 LJ'' potential) and those of sv30 
for the ``with hard-core repulsion'' 
potential. That the approximate exponents $0.39$ and $0.19$ in the above 
scaling relations are both significantly smaller than unity implies 
that explicit-chain simulated LLPS properties with background LJ, at 
least as far as $T^*_{\rm cr}$ and $\phi_{\rm cr}$ are concerned, 
are less sensitive to sequence charge pattern than that predicted 
by RPA+FH theories. However, our finding that sv30 (an outlier in Fig.~14a) 
can---but sv1 and sv15 cannot---phase separate in the ``with hard-core 
repulsion'' model (Fig.~7) suggests that in this case LLPS in 
explicit-chain models can be even 
more sensitive to sequence charge pattern than that in RPA.
In other words, our results suggest that for polyampholytes that 
interact only via electrostatics 
and hard-core excluded-volume repulsion, pure RPA can overstimate 
LLPS propensity. The case in point here is that while
RPA predicts LLPS for sv1 and 
sv15 with $T^*_{\rm cr}$'s that are 0.0104 and 0.149, respectively, of
the $T^*_{\rm cr}$ of the sv30 sequence,\cite{lin2017,suman1} 
the sv1 and sv15 sequences do not phase separate at temperature
much lower---as low as $0.001/1.65=6.06\times 10^{-4}$ that of sv30's 
$T^*_{\rm cr}$---when simulated using the explicit-chain model in Fig.~7.

Figure~14b and Table~2 show that the simulated $\phi_{\rm cr}$'s
are larger than their theoretical counterparts for the systems we studied.
However, the range of variation is much smaller for the simulated
$\phi_{\rm cr}$'s (from $0.082$ to $0.152$) than for the theoretical
$\phi_{\rm cr}$'s (from $0.0123$ to $0.124$). It follows that there
is a substantial variation in the ratio $\phi_{\rm cr, sim}/\phi_{\rm cr,thr}$
of simulated to theoretical critical volume fraction, from
$0.133/0.124=1.07$ for sv1 in the ``with 1/3 LJ'' model to
$0.082/0.0123=6.67$ for sv30 in the ``with hard-core repulsion'' model. 
For sequences sv1, sv15, and sv30 simulated using the same interaction scheme, 
this ratio increases from sv1 to sv15 to sv30, as manifested already by the 
approximate $\phi_{\rm cr,sim}\sim$ $(\phi_{\rm cr,thr})^{0.19}$ scaling
noted above. The large $\phi_{\rm cr, sim}/\phi_{\rm cr,thr}$ ratio for 
sv30 in the ``with hard-core repulsion'' model is quantitatively in line 
with our previous comparison of lattice-simulated and RPA-predicted 
$\phi_{\rm cr}$'s (Figure~12c of ref\cite{suman1}).
In contrast, the smaller $\phi_{\rm cr, sim}/\phi_{\rm cr,thr}$ ratios
likely arise from the FH contribution to some of the present
theoretical $\phi_{\rm cr}$'s.
Pure FH predicts $\phi_{\rm cr}=(\sqrt{N}+1)^{-1}$ [critical 
$\chi_{\rm cr}=(\sqrt{N}+1)^2/(2N)$]. For the current systems with $N=50$,
this formula translates to $\phi_{\rm cr}=0.124$, which tends to be significant 
larger than that predicted by pure RPA.

Echoing the observation that the as1, as2, as3, and as4
sequences are outliers with regard to RPA-predicted properties 
(Fig.~3c and Fig.~12), these sequences are also outliers in 
Fig.~14. If linear regression is applied in Fig.~14 to these sequences
alone, the correlation coefficient for $T^*_{\rm cr}$ in Fig.~14a becomes
$r=-0.868$, with a regression slope $-0.204$ that is opposite in sign
to that for all the plotted data points (slope $=+0.387$).  No clear
trend is discernible for these four sequences in the theory-simulation
comparison of $\phi_{\rm cr}$ in Fig.~14b ($r=0.198$).
(Removing the data points for these four sequences has only 
very limited effects on the overall linear regressions for Fig.~14a and
for Fig.~14b).
As emphasized above, the peculiar theoretical and simulated LLPS 
properties of the as1--4 sequences as well as how these properties 
are governed by their charge patterns deserve further examination. 
\\

\noindent
{\Large\bf 5 Conclusions}\\

In summary, we have taken a step to improve the currently limited 
understanding of the sequence-dependent physical interactions that 
underlie LLPS of IDPs by extensive simulations 
of explicit-chain models that allow for a coarse-grained
representation of IDP at the residue level, using multiple-chain
systems each consisting of 500 individual chains.
By analyzing results for 50-residue sequences with diverse
charge patterns using model interaction potentials 
consisting of different combinations of sequence-dependent electrostatics,
hard-core excluded-volume repulsion, and LJ attractions, 
we find that while a general inter-residue LJ attraction---which has a short
spatial range---favors LLPS, such a background short-range attraction 
diminishes sequence specificity of LLPS. Interestingly, and consistent with 
RPA theory, the condensed side of the coexistence curve of one of the
polyampholytes we simulated exhibits a pronounced upward concavity in 
the absence of background LJ attraction.
Such upward concavity is not observed in the presence of strong 
background LJ interaction or in classical FH theory.
This finding suggests that long-range electrostatic repulsion
likely allows for condensed phases that are more dilute than when 
short-range attraction is prominent. This observation should
contribute insights into the physical forces that maintain 
condensed-phase volume fractions of $\approx 0.2$ or even 
lower.\cite{jacob2017,low-rho} It should be relevant as well for future 
development of computational and theoretical studies of IDP LLPS that
address other sequence-dependent energies \cite{biochemrev,robert} beyond 
electrostatics and LJ-like hydrophobic interactions.\cite{dignon18,suman1}

A main goal of the present study is to use explicit-chain simulations
to assess the accuracy of analytical theories and the utility of simple 
sequence charge pattern parameters $\kappa$ and SCD in capturing LLPS 
properties of IDPs.  The calculation of a pattern parameter for a sequence is
virtually instantaneous and numerical calculations for analytical theories
are far less computationally intensive than explicit-chain simulations.
Therefore, in addition to being tools for elucidating LLPS physics,
sequence charge pattern parameters and analytical theories can contribute
to efficient high-throughput bioinformatics studies and the screening of 
candidates in IDP sequence design. Here we have compared and 
contrasted results simulated for several polyampholyte sequences 
using the present explicit-chain model against the corresponding 
analytical theory predictions. A broader context for this evaluation is
provided by RPA-predicted critical temperatures and volume fractions 
we calculated for 10,000 randomly sampled sequences.
For three sequences belonging to a previous studied set, the simulated
critical temperatures, $T^*_{\rm cr}$'s, correlate reasonably well with 
theoretical predictions and also with the $\kappa$ and SCD parameters. 
We find that the simulated $T^*_{\rm cr}$'s are less sensitive 
to sequence charge pattern than their theory-predicted counterparts 
when a substantial background LJ interaction is in play. However, simulated
$T^*_{\rm cr}$'s can be more sensitive than RPA-predicted $T^*_{\rm cr}$'s
in the absence of background LJ interaction. In this regard,
our results suggest that LLPS propensity can be overestimated by RPA 
in such cases for sequences with small $\kappa$ and small $-$SCD values.
Most notably, for four sequences intentionally generated as outliers
in the $\kappa$-SCD relationship, neither $\kappa$ nor SCD is
a LLPS predictor with a reliable discriminatory power. This discovery
suggests that the effect of blockiness of sequence-local charge pattern on 
LLPS may be overestimated by $\kappa$, whereas the nonlocal effect
of sequence charge pattern on LLPS may be overestimated by SCD.
Therefore, a more generally applicable sequence charge pattern parameter 
for LLPS propensity should be developed to overcome this limitation.
All in all, in view of the new questions posed by our findings, there is 
no shortage of productive avenues of further investigation into the physical 
basis of biomolecular condensates.
\\

$\null$\\
{\large\bf Conflicts of Interest}\\ 
There are no conflicts of interest to declare.

$\null$\\
{\large\bf Acknowledgments}\\ 
We thank Julie Forman-Kay for helpful discussions.
This work was supported by Canadian Institutes of Health Research grants
MOP-84281 and NJT-155930, Natural Sciences and Engineering Research Council
of Canada Discovery Grant RGPIN-2018-04351, and computational 
resources provided by SciNet of Compute/Calcul Canada.

\vfill\eject 
\noindent
{\Large\bf References}\\
 
 \vfill\eject

{{\bf Table 1.} Charge pattern parameters for the
sequences studied in this work.}
\begin{center}
\begin{tabular}{ccc} 
\hline 
$\;$ sequence $\;$ & $\quad$ SCD $\quad$ & $\quad$ $\kappa$ $\quad$\\
 \hline 
sv1 & $-0.413$ & $0.0009$ \\
sv15 & $-4.349$ & $0.1354$ \\
sv30 & $-27.84$ & $1.0000$ \\
as1 & $-12.79$ & $0.1761$ \\
as2 & $-10.30$ & $0.4853$ \\ 
as3 & $-8.266$ & $0.6125$ \\
as4 & $-6.11$ & $0.7783$ \\
\hline
\end{tabular}  
\end{center}
 \vfill\eject

{{\bf Table 2.} Simulated and theoretical critical temperatures
and critical densities or volume fractions considered in Fig.~12 
and Fig.~13. Data for sequences sv1 and sv15 in the lattice model 
and their theoretical counterparts (last two rows) are obtained from 
Das et al.\cite{suman1} Other data are from the present study.
}
\begin{center}
\begin{tabular}{|c|c||c|c||c|c|}
\hline 
Potential type & Sequence & \multicolumn{2}{|c||}{Simulation} & 
\multicolumn{2}{|c|}{Theory} \\
\cline{3-6}
$\null$ & $\null$ & $T^*_{\rm cr}$ & $\rho_{\rm cr}$ & 
$T^*_{\rm cr}$ & $\phi_{\rm cr}$ \\
\hline
``with LJ'' & sv1 & 3.52 & 0.152 & 4.55 & 0.124 \\
\cline{3-6}
            & sv15 & 3.86 & 0.130 & 4.93 & 0.098 \\
\cline{3-6}
            & sv30 & 4.97 & 0.120 & 10.52 & 0.019 \\
\hline
``with 1/3 LJ'' & sv1 & 1.20 & 0.133 & 1.52 & 0.124 \\
\cline{3-6}
                & sv15 & 1.52 & 0.127 & 2.14 & 0.040\\
\cline{3-6}
                & sv30 & 3.44 & 0.086 & 9.14 & 0.014 \\
\cline{3-6}
                & as1 & 2.25 & 0.095 & 4.43 & 0.017\\
\cline{3-6}
                & as2 & 2.31 & 0.096 & 3.77 & 0.020\\
\cline{3-6}
                & as3 & 2.28 & 0.110 & 3.63 & 0.025\\
\cline{3-6}
                & as4 & 2.41 & 0.095 & 3.27 & 0.032\\
\hline
``with hard-core & sv30 & 1.65 & 0.082 & 8.57 & 0.0123\\
  repulsion''    &    &      &       &   & \\
\hline
screened        & sv1 & 0.70 & 0.106$^{\rm a}$ & 0.114 & 0.0486\\
\cline{3-6}
                & sv15 & 1.34 & 0.105$^{\rm a}$ & 1.091 & 0.0187\\
\hline
\multicolumn{6}{l}
{$\null^{\rm a}$This quantity equals the simulated critical volume}\\
\multicolumn{6}{l}
{{\phantom{$\null^{\rm a}$}}fraction $\phi_{\rm cr}$ in the 
lattice model.\cite{suman1}}\\
\end{tabular}  
\end{center}
 \vfill\eject

\begin{figure}[t]
   \centering
   \includegraphics[width=0.98\columnwidth]{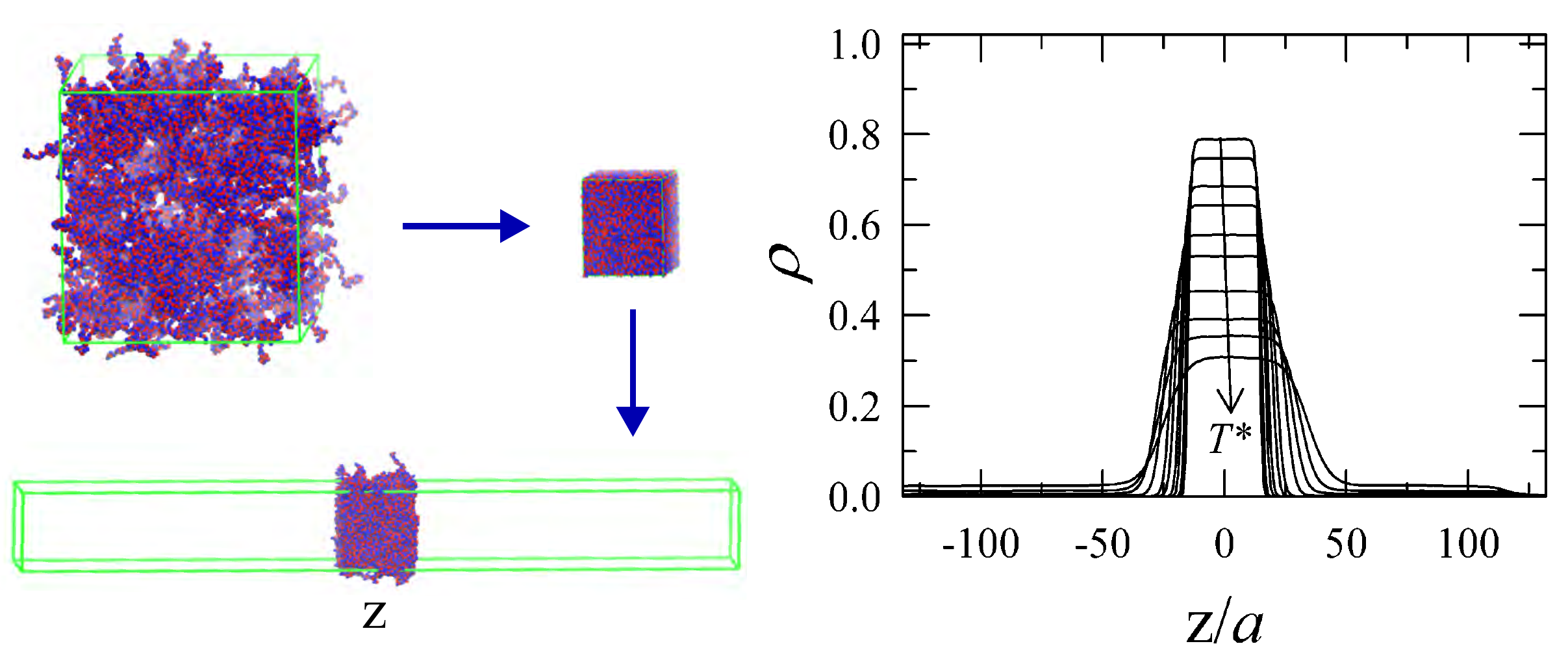}
   \caption{
Simulation methodology. The schematic ({\it left}) illustrates the
computational technique \cite{dignon18,panag2017} we adopt for
calculating phase diagrams of polyampholytes. After energy minimization
of a collection of model polyampholytes (chains of red and blue beads)
at $T^*=4.0$, the cubic simulation box (green frame)
with periodic boundary conditions (visualized using 
VMD \cite{vmd} with chains at boundaries unwrapped) 
is compressed under the same high temperature 
(blue horizontal arrow). This is followed by an expansion 
(blue vertical arrow) of the simulation box at 
$T^*=1.0$ along the direction (labeled by 
Cartesian coordinate $z$) of one of its edges,
resulting in an enlarged simulation box taking the shape of an elongated 
rectangular cuboid with a slab of polyampholytes centered at $z=0$.
The system is then equilibrated at different temperatures. The plot 
({\it right}) shows an example of temperature-sensitive equilibrated
distributions of polyampholyte density $\rho$ as 
a function of $z$ (in units of $a$) for the strictly alternating sequence
sv1 (refs\cite{pappu13,suman1}, see below). The downward pointing
arrow indicates increasing $T^*$. For a given temperature,
the maximum of the distribution is identified as the polyampholyte
density of the condensed phase whereas the minimum
as the polyampholyte density of the dilute phase. A phase diagram can 
thus be constructed from these data. See text for further details. 
}
   \label{fig1}
\end{figure}

$\null$\\

\vfill\eject

\begin{figure}[t]
   \centering
   \includegraphics[width=0.68\columnwidth]{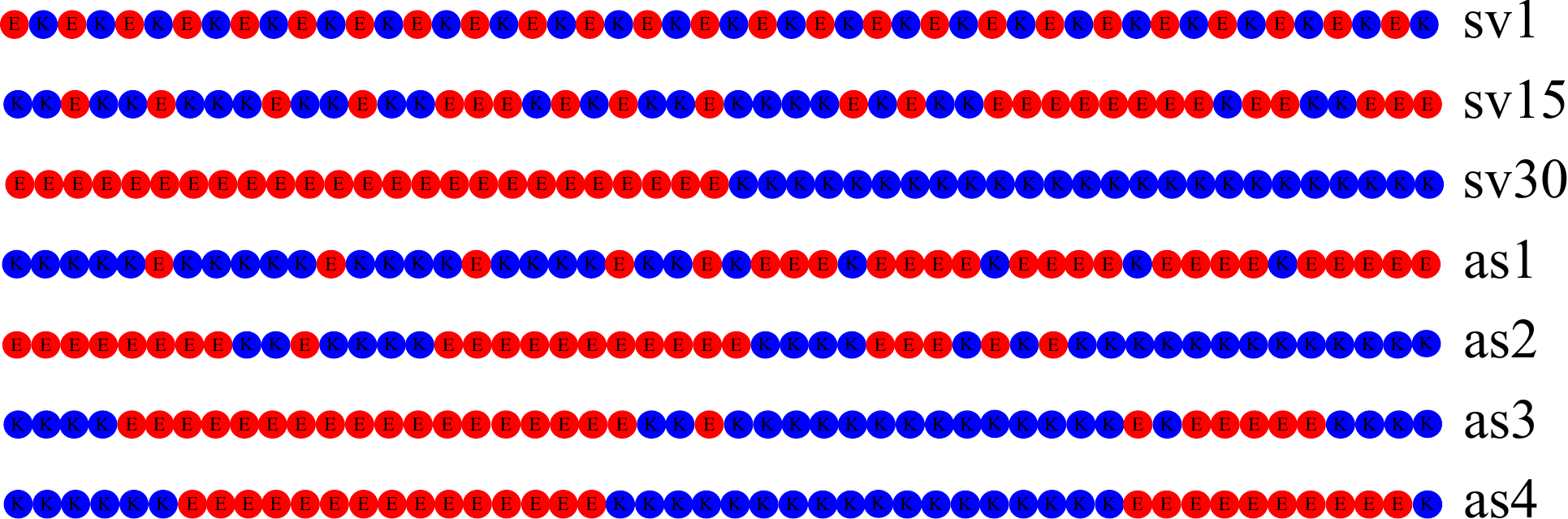}
   \caption{
The polyampholyte sequences studied in this work. Every sequence contains 25
K's (blue beads) and 25 E's (red beads) with different arrangements 
of K's and E's along the sequences. Sequences sv1, sv15, and sv30 are 
from ref\cite{pappu13}; sequences as1, as2, as3, and as4 are introduced 
by the present
work.
}
   \label{fig2}
\end{figure}

$\null$\\

\vfill\eject

\begin{figure}
   \centering
   \includegraphics[width=0.43\columnwidth]{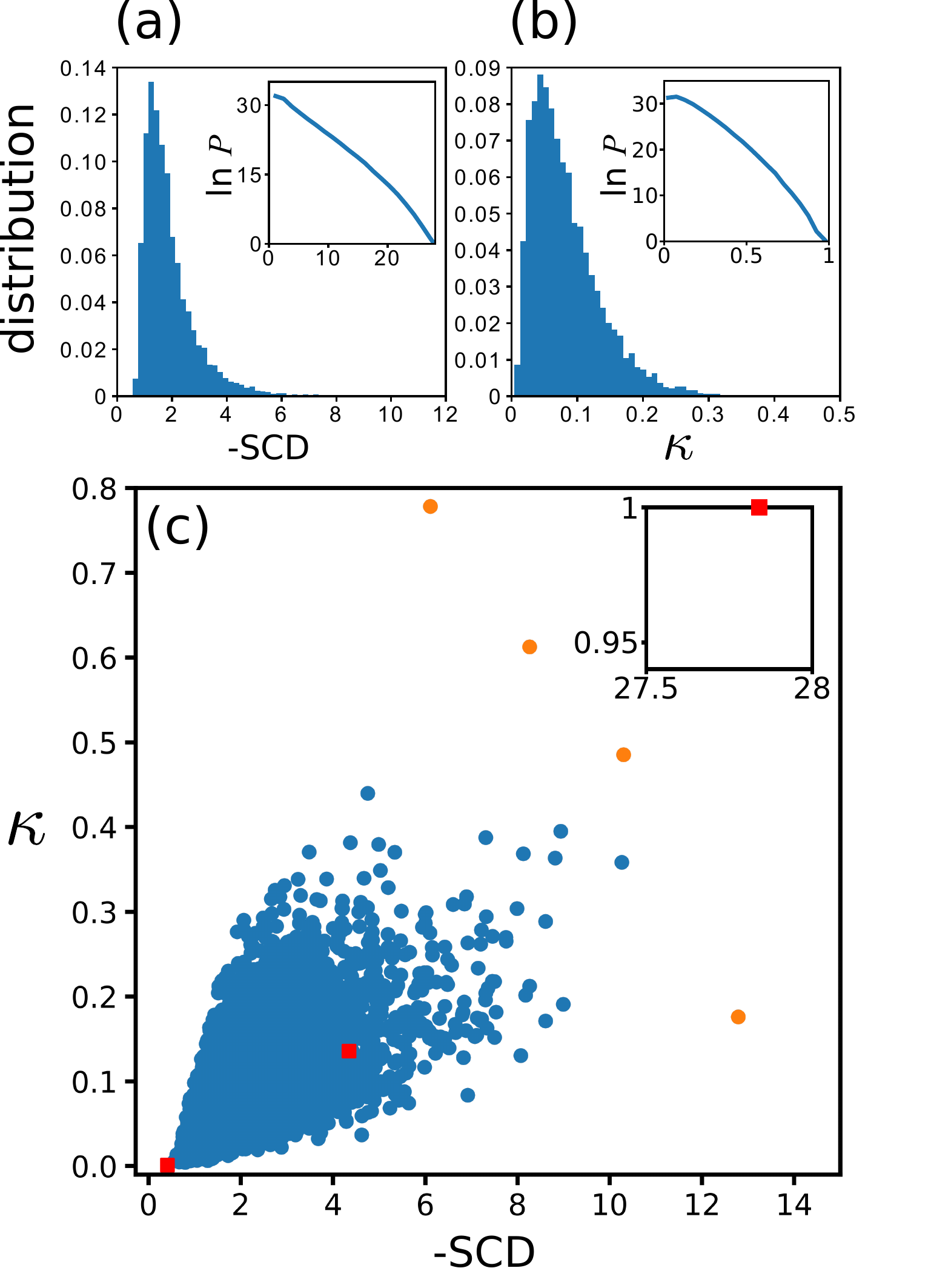}
   \caption{
Sequence-space statistics of charge pattern parameters. Normalized 
distributions of (a) SCD [Eq.~(5)] and (b) $\kappa$ [Eq.~(4)] 
among 50-residue fully charged 
but overall neutral KE sequences are computed from 10,000 sequences 
generated by repeat exchanges of sequence positions of randomly selected
pairs of positive (K) and negative (E) residues. These 
distributions---histograms in (a,b)---cover only the readily sampled range of 
SCD and $\kappa$ values, whereas the full distributions of sequence 
population $P$ over the entire range of all possible SCD and $\kappa$ values 
are estimated using the Wang-Landau technique \cite{WL1,WL2} 
[semi-log plots in the insets of (a) and (b)].  (c) The $-$SCD 
and $\kappa$ values of the sv1, sv15, sv30 sequences (red squares,
{\it bottom to top}) and the as1, as2, as3, and as4 sequences (orange circles,
{\it bottom to top}) are shown against the backdrop
of the $-$SCD versus $\kappa$ scatter plot for the 10,000 randomly sampled
sequences (blue circles). Sequence sv30 is shown in the
inset of (c) because it lies outside the range of the scatter plot.
In the histograms in (a) and (b), each of the horizontal ranges between 
$-$SCD $=0.5640$ and $10.2632$ (a) and that between $\kappa=0.0044$ and 
$0.4396$ (b) is equally divided into 50 bins. The height of each bar
in the histograms is a normalized bin population that is inclusive of 
the lower boundary but exclusive of the upper boundary of the given bin
except it is inclusive of both boundaries for the bin with the
largest $-$SCD or $\kappa$.  The insets
of (a) and (b) are obtained by averaging over ten Wang-Landau processes 
initialized by different sequences; sampled sequences 
are binned into 20 equal intervals for the full range of SCD and 
$\kappa$ values using the same rule for inclusion/exclusion of bin 
boundaries as described above for the other bins over more limited SCD
and $\kappa$ ranges. The scale for population $P$ is 
such that the sum of all 20 binned populations is equal to the total 
number, $50!/(25!25!)$, of fully charged 50-residues KE sequences with 
zero net charge. For the same reason, $P$ is
set to unity for the maximum value of $-$SCD and for $\kappa=1$ since 
both of these parameter values uniquely specify the diblock sv30 sequence. 
Assuringly, the trend in the inset of (b) is very similar to
that exhibited by the previously estimated population distribution 
over $\kappa$ in Fig S1 of ref\cite{pappu13}. 
}
   \label{fig3}
\end{figure}

\vfill\eject

\begin{figure}[t]
   \centering
   \includegraphics[width=\columnwidth]{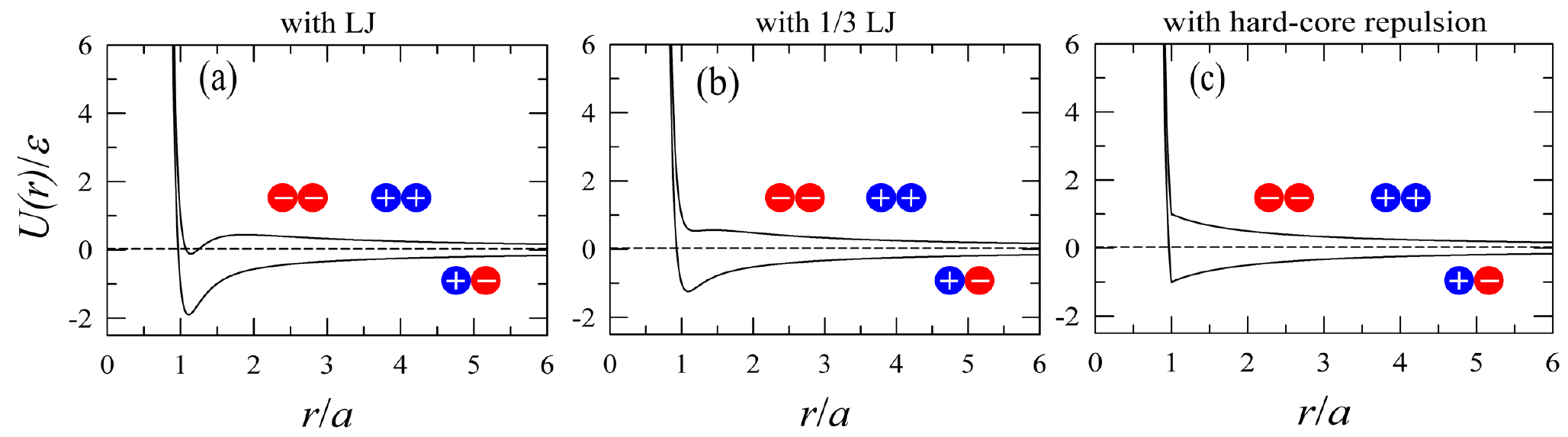}
   \caption{
Inter-residue interaction potentials used in our explicit-chain models.
In each panel, total interaction energy $U(r)$ in units of $\epsilon$ is 
shown as a function of inter-residue distance $r$. The upper and lower 
curves are for a pair of interacting residues with like and opposite 
charges, respectively, as illustrated by the red- and blue-bead 
representations of charged residues. 
The $U(r)=0$ level is marked by a horizontal dotted line.
(a) Electrostatics + LJ model [Eq.~(1) plus Eq.~(2)]. 
(b) Electrostatics + 1/3 LJ model [Eq.~(1) plus 1/3 of Eq.~(2)].
(c) Electrostatics + hard-core repulsion model
[Eq.~(1) plus a modified form of Eq.~(2) for which the entire LJ term
is set to zero for $r>a$]. 
}
   \label{fig4}
\end{figure}

\vfill\eject

\begin{figure}
   \centering
   \includegraphics[width=\columnwidth]{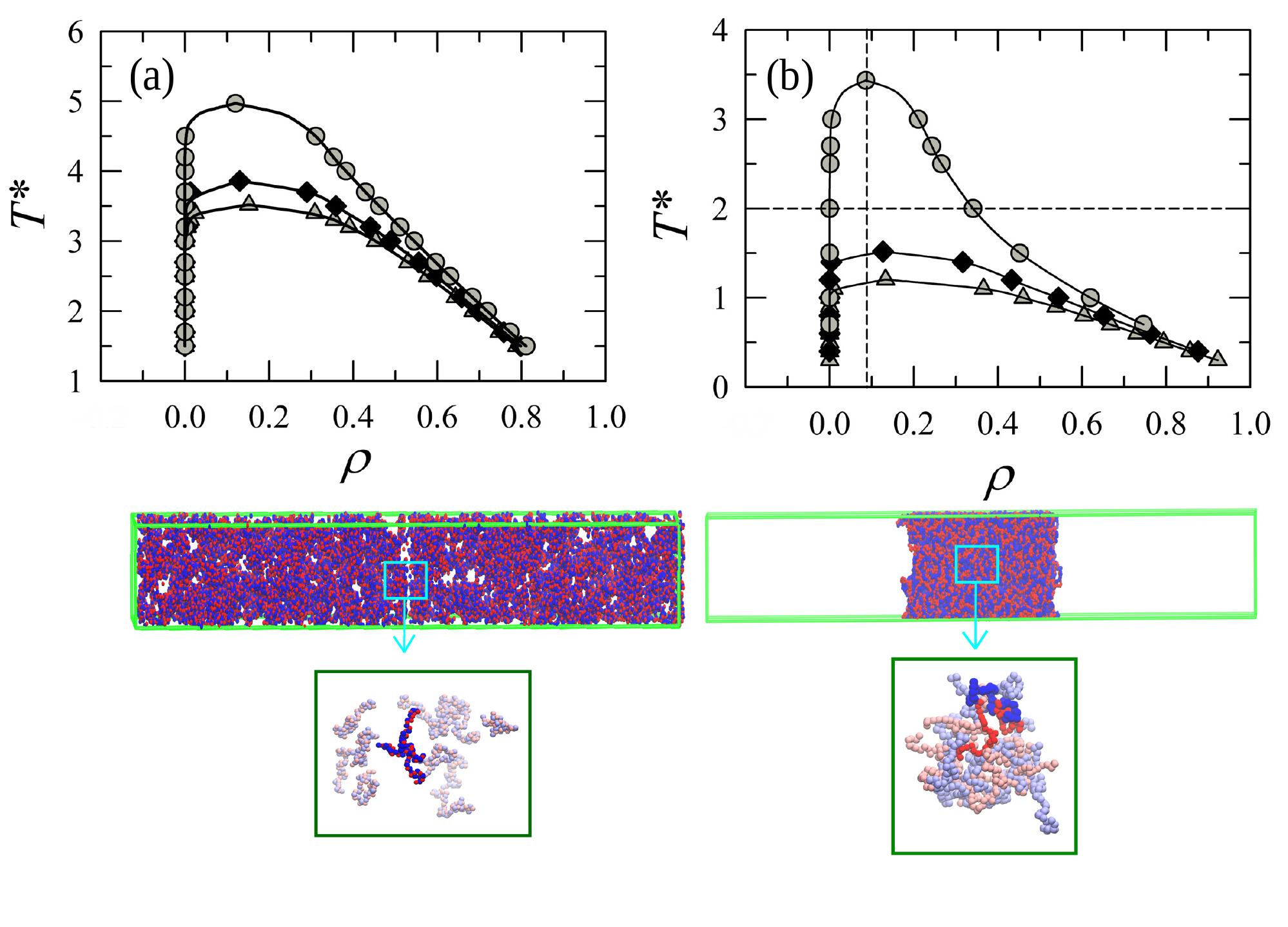}
   \caption{
Charge-pattern-dependent phase separations. Phase diagrams are calculated
for sequences sv1 (triangles), sv15 (diamonds), and sv30 (circles) using
(a) the electrostatics + LJ model potential (Fig.~4a) and (b) the
electrostatics + 1/3 LJ model potential (Fig.~4b). Fitted coexistence curves 
simulated using the present coarse-grained continuum explicit-chain model 
here and in subsequent figures are constructed as described \cite{panag2017} 
and serve largely as guides to the eye.  In (b), the vertical 
dashed line marks the critical density of sv30 to underscore that it
is lower than the critical densities of sv1 and sv15. The horizontal
dashed line marks the $T^*=2.0$ for which one snapshot of each of the 
simulation boxes (green frames) for sequence sv1 ({\it left}) and for 
sequence sv30 ({\it right}) in (b) are shown below the phase diagrams.
Renditions of close-up images (dark-green boxes, {\it bottom}) of selected 
parts of the simulation boxes (blue boxes with arrows) are 
provided to illustrate key differences in local chain configuration between 
the two systems. Each of the images in the bottom dark-green boxes consists
of one randomly chosen chain and every chain that has either
5 or more ({\it left} for sv1) or 15 or more ({\it right} for sv30)
residues positioned within a distance of $6a$ from a residue of the 
chosen chain. A pair of such chains are depicted in more saturated
color for the sole purpose of enhancing the visual effect.
}
   \label{fig5}
\end{figure}

\vfill\eject

\begin{figure}
   \centering
   \includegraphics[width=\columnwidth]{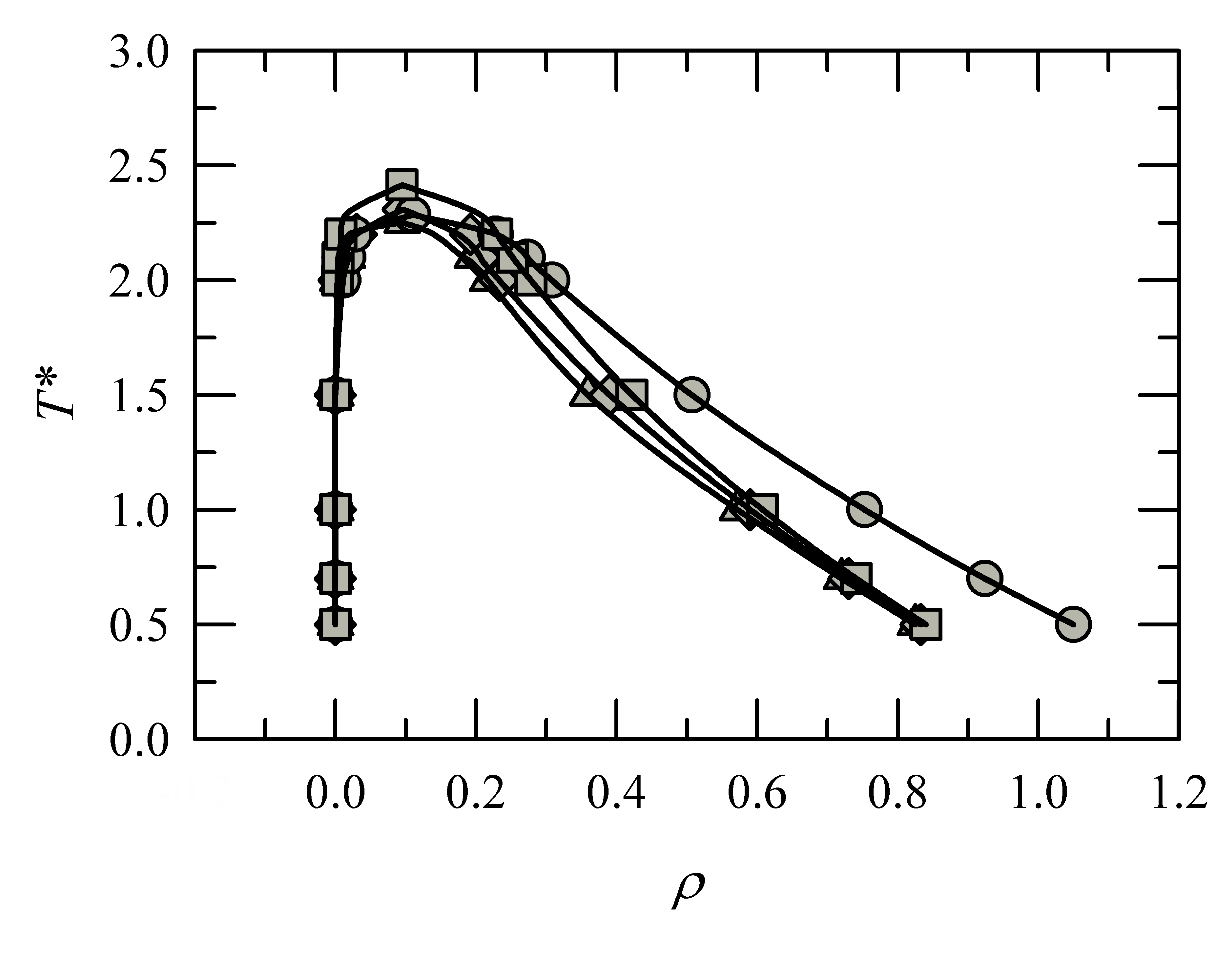}
   \caption{
Phase diagrams for the four newly introduced polyampholyte sequences.
Simulation results are shown for sequences as1 (triangles), as2 (diamonds), 
as3 (circles), and as4 (squares), all computed using the electrostatics 
+ 1/3 LJ potential (Fig.~4b). Critical temperature and critical volume are 
quite insensitive to the variation of charge pattern among these sequences. 
}
   \label{fig6}
\end{figure}

\vfill\eject

\begin{figure}
   \centering
   \includegraphics[width=\columnwidth]{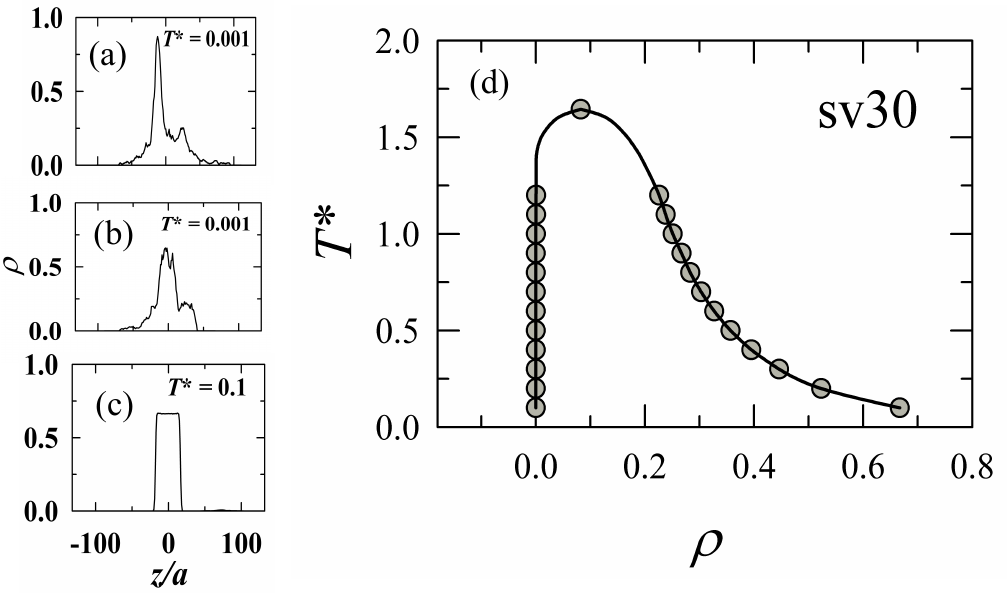}
   \caption{
Phase behaviors in the ``with hard-core repulsion'' model.
Polyampholyte density as a function of $z$ is calculated using 
the model potential in Fig.~4c
for (a) sv1, (b) sv15, and (c) sv30 at the temperatures
indicated. (d) Phase diagram for sequence sv30 in the same model.
}
   \label{fig7}
\end{figure}

\vfill\eject

\begin{figure}
   \centering
   \includegraphics[width=\columnwidth]{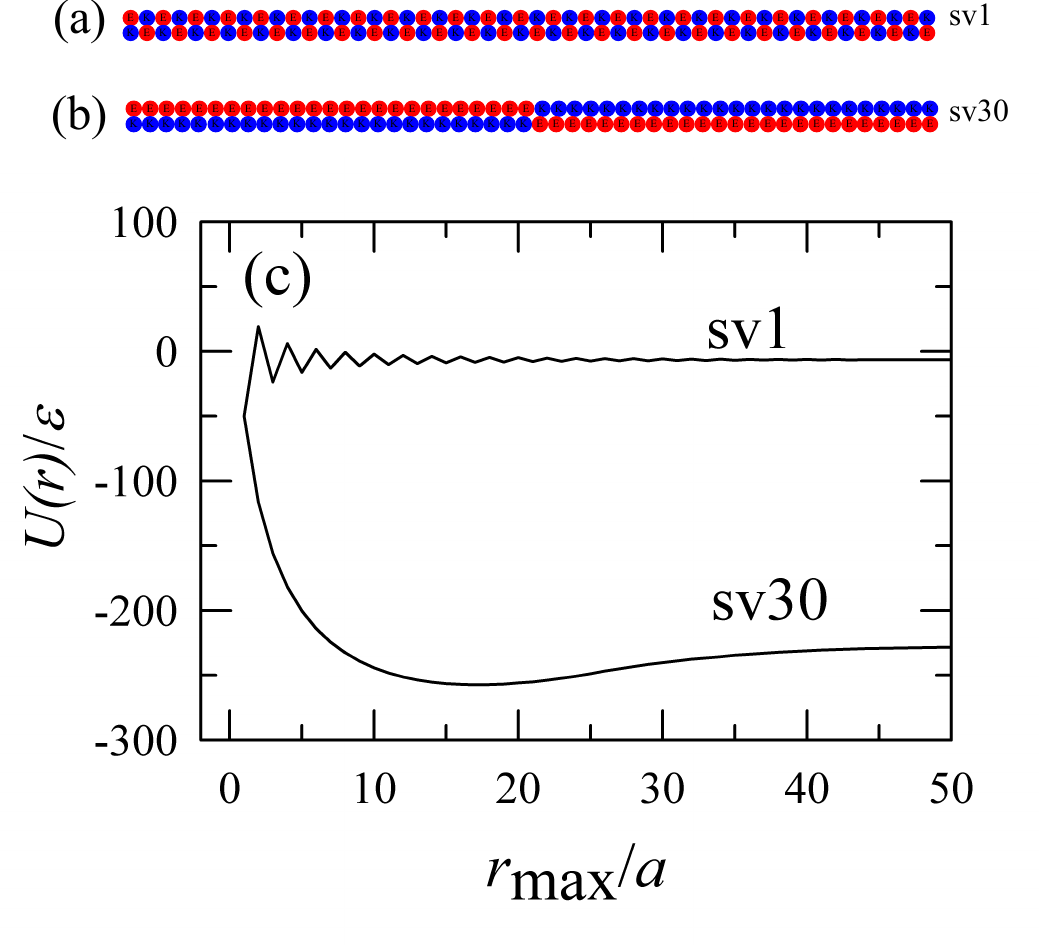}
   \caption{
Inter-chain polyampholyte interactions are strongly sequence dependent.
A pair of sv1 sequences (a) and a pair of sv30 sequences are each shown
in an energetically favorable aligned configuration.
(c) Total interaction energy for the configurations in (a) and (b) as
functions of the maximum residue-residue distance, $r_{\rm max}$,
that is taken into consideration in computing the electrostatic
energies.
}
   \label{fig8}
\end{figure}

\vfill\eject

\begin{figure}
   \centering
   \includegraphics[width=\columnwidth]{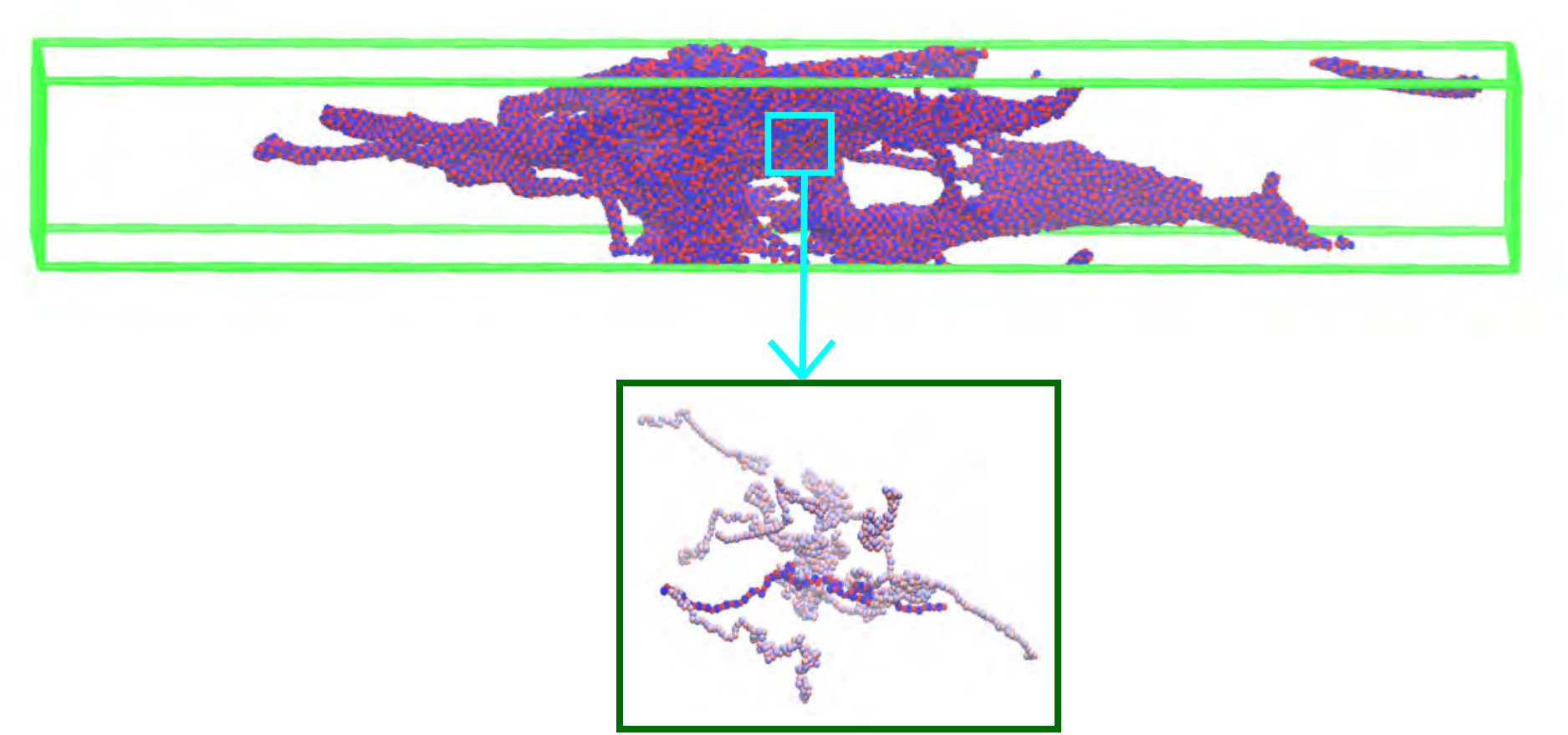}
   \caption{
Strictly alternating polyampholytes with hard-core repulsion at
extremely low temperature.
A snapshot of sv1 chains in a simulation box at $T^*=0.01$. The chains
are seen as associated and not scattered though there is no clear sign 
of phase separation (cf. Fig.~7a). The close-up image (bottom) 
includes all of the chains that have at least one 
residue within $6a$ of a residue on a randomly selected chain. 
(The actual number of such $\le 6a$ 
inter-chain residue-residue distances varies from 1 to 24 in the chain 
cluster shown). Two chains inside the bottom box are depicted in more 
saturated color for the sole purpose of enhancing visualization.
}
   \label{fig9}
\end{figure}

\vfill\eject

\begin{figure}
   \centering
   \includegraphics[width=0.7\columnwidth]{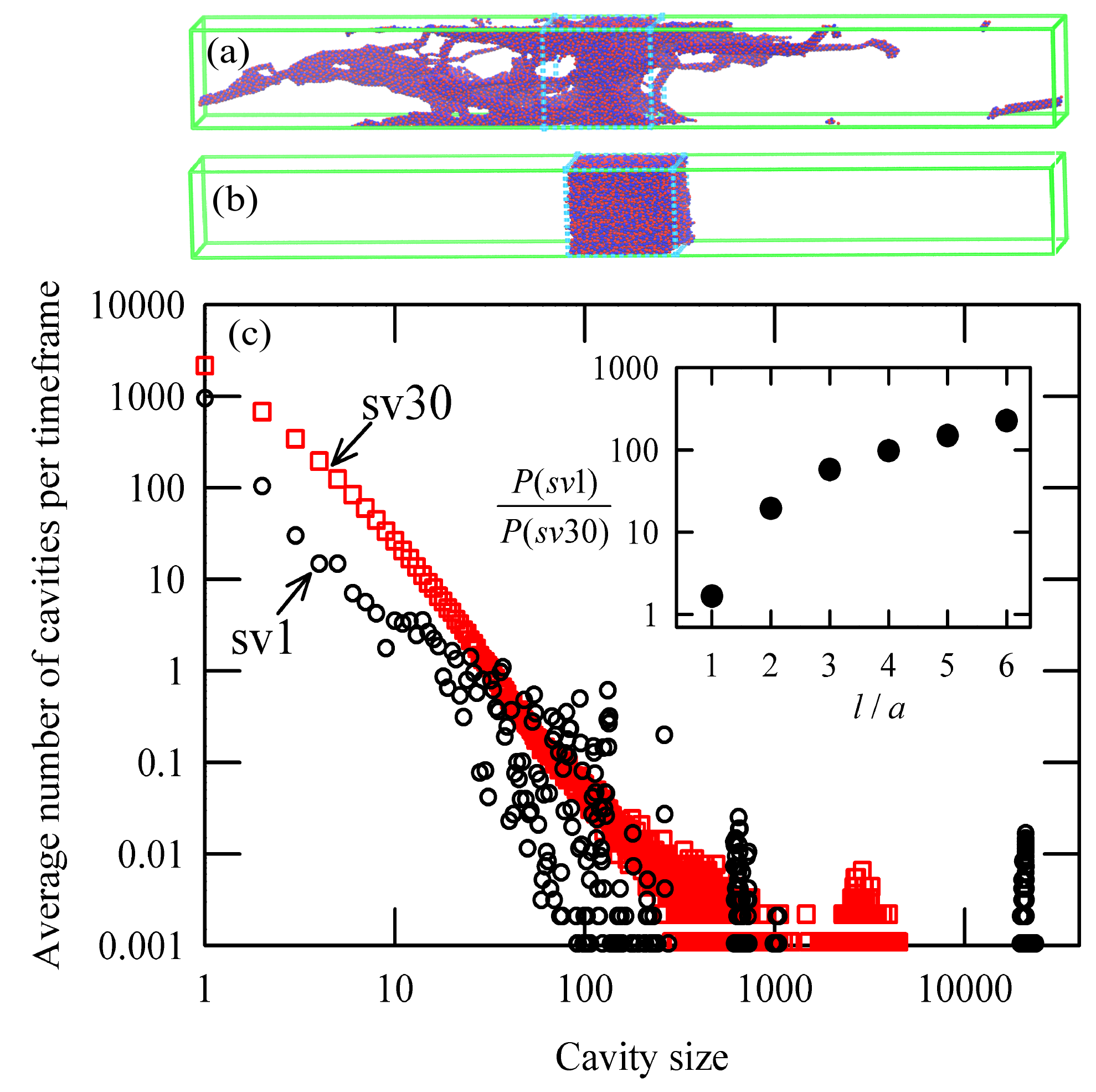}
   \caption{
Comparing distributions of cavity size in different polyampholyte-rich states.
The associated state of sv1 at $T^*=0.001$ 
in the ``electrostatic + hard-core repulsion'' model (a) and
the condensed phase of sv30 at $T^*=0.7$ in the
``electrostatics + 1/3 LJ'' model (b) are compared by considering
959 and 914 configurational snapshots (timeframes), 
respectively, for sv1 and sv30 
[one snapshot of each set is shown in (a) and (b)]. We focus 
on their respective volumes of $(33a)^3$ with highest average polyampholyte
density [indicated by dotted light-blue boxes in (a,b); $z\in [-29a,4a]$ 
in (a) and $z\in [-16.5a,16.5a]$ in (b)], with periodic boundary
conditions maintained within these volumes of $(33a)^3$ along $x$ and $y$ 
but not in the $z$ direction. We consider small cubic 
volumes of $l^3$, where $l=a,2a,\dots,6a$,  
that are placed at intervals of $a$ in all three spatial directions
within these volumes and determine the number of such small cubic
volumes that do not contain the center of any of the monomers that
make up the polyampholytes. These small cubic volumes are termed
empty. The configurational difference between the polyampholyte-rich
states of the sv1 and sv30 systems here is characterized by two measures.
First, a cavity is identified as a region covering one $l=a$ empty volume 
(of $a^3$) and all $l=a$ empty volumes contiguous to it directly or indirectly
(i.e., a given $l=a$ empty volume can only belong to one cavity).
The size of cavity is given by the number of contiguous $l=a$ empty volumes. 
The distribution of cavity volume so defined is given in (c) for
sv1 (black open circles) and sv30 (red open squares). 
Second, the total number, $P$, of positions of empty volumes for various 
$l$ are counted, and the ratio of $P$ for sv1 to that for sv30 is taken
for various $l$ values. The inset in (c) shows 
$P({\rm sv1})/P({\rm sv30})>1$ and increases sharply with increasing $l$.
Hence both measures indicate that there are substantially
more cavities of larger volumes for the sv1 system 
than for the sv30 system analyzed in this figure.
}
   \label{fig10}
\end{figure}

\vfill\eject

\begin{figure}
   \centering
   \includegraphics[width=\columnwidth]{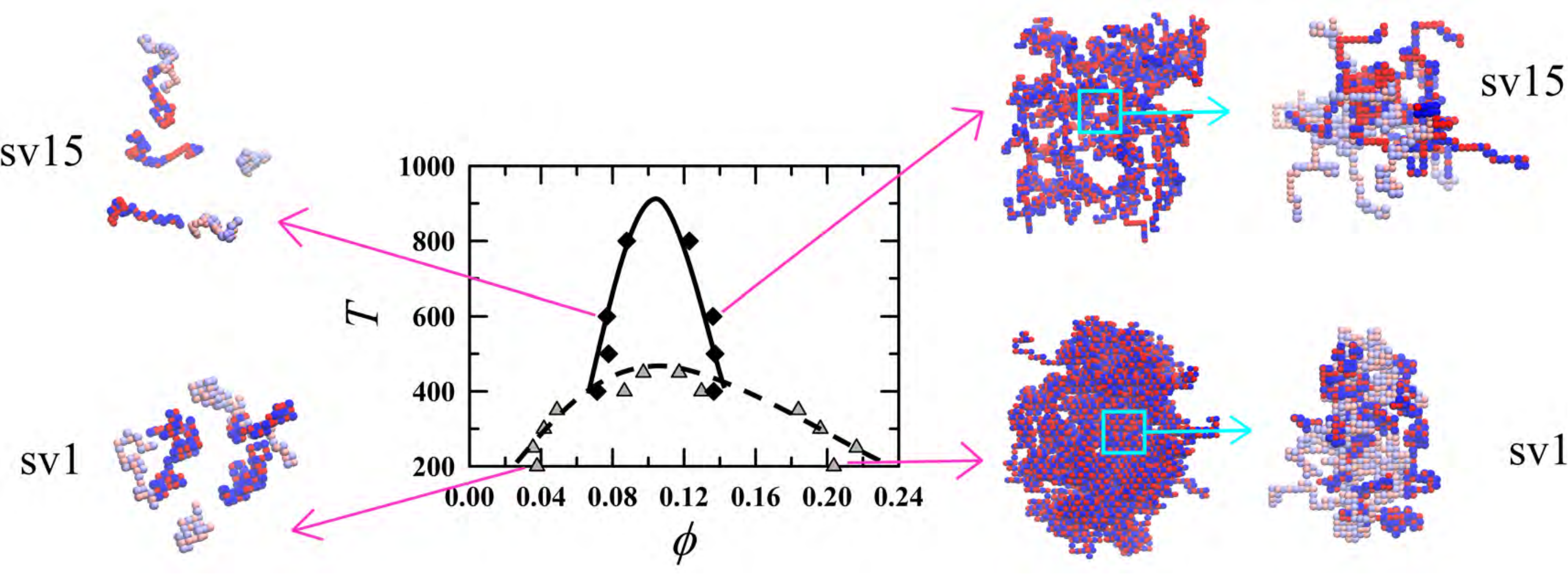}
   \caption{
Lattice chain configurations in dilute and condensed phases of phase-separated
polyampholytes. The phase diagrams (center) for sv1 (triangles fitted
by dashed curve) and for sv15 (diamonds fitted by solid curve)  were 
computed using our previous lattice model and adapted using data from 
Figure~8 of Das et al.\cite{suman1} $T$ is absolute temperature
and $\phi$ is polyampholyte volume fraction as described.\cite{suman1}
Here we provide snapshots of the dilute and of the condensed phases
in the lattice model under the simulated $(T,\phi)$ conditions 
indicated by the red arrows.
Snapshots on the dilute side consist of chains in a randomly selected
volume within the low-$\phi$ region of the simulation box. 
Snapshots on the condensed side show substantial fractions of the
condensed phases as well as close-up images of parts of them
(marked by light boxes with arrows). The close-up images are rendered using
the same protocol as that for the bottom-right image in Fig.~5.
Selected chains in the snapshots of the dilute state and in the close-up
images are depicted in more saturated colors than others for the sole 
purpose of enhancing visualization.
}
   \label{fig11}
\end{figure}

\vfill\eject

\begin{figure}
   \centering
   \includegraphics[width=\columnwidth]{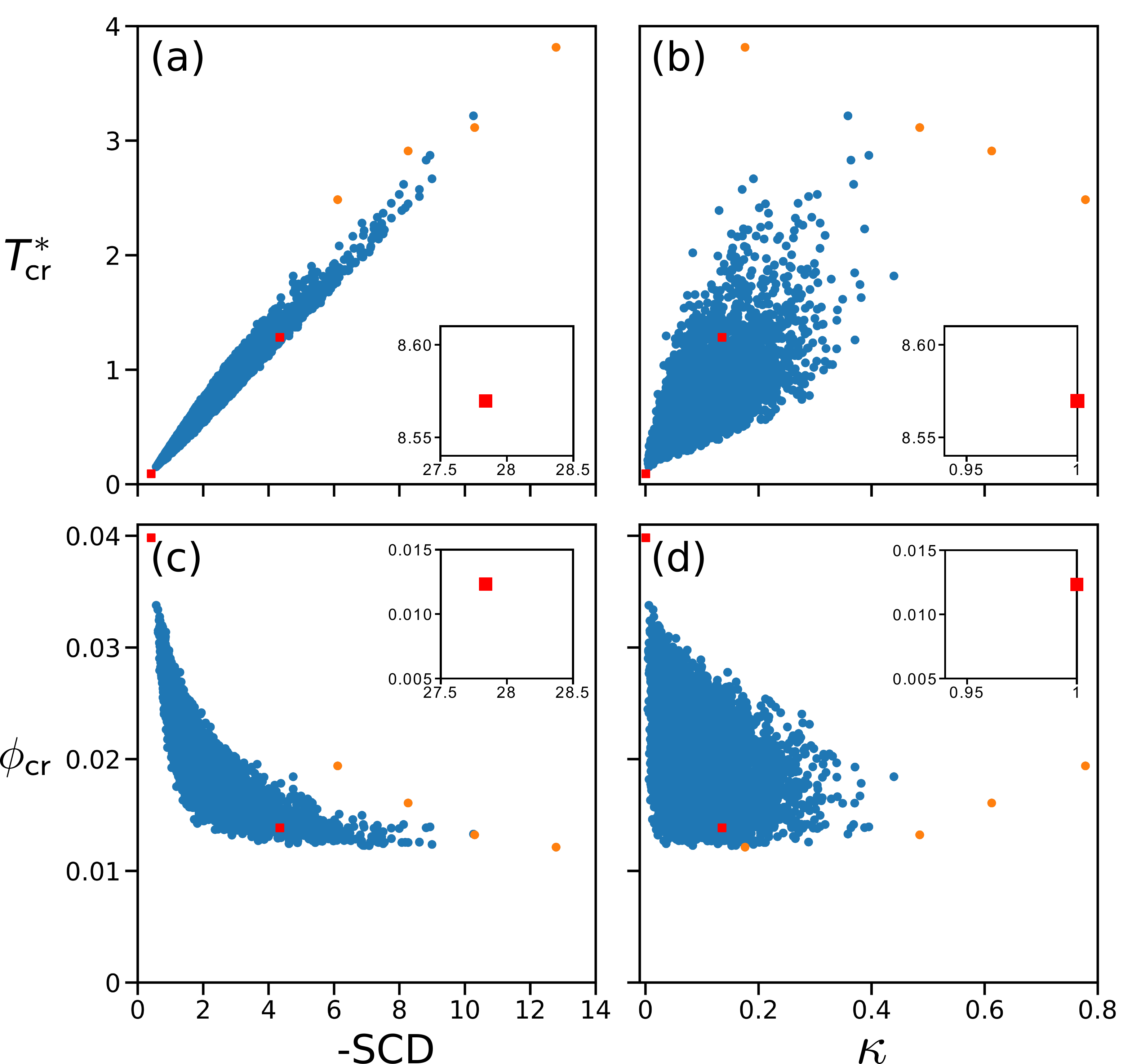}
   \caption{
Relationship between RPA-predicted phase properties and sequence charge
pattern parameters.
RPA-predicted critical temperatures $T^*_{\rm cr}$ (a,b) and critical volume 
fractions $\phi_{\rm cr}$ (c,d) of the 10,000 sampled polyampholyte sequences 
in Fig.~3 are computed using the salt-free RPA formulation\cite{linPRL}
(no Flory $\chi$ parameter) and plotted against their $-$SCD (a,c) and 
$\kappa$ (b,d) values (blue circles).
The sequences studied by the present explicit-chain simulations
are marked as in Fig.~3 (red squares for sv1, sv15, and sv30; orange
circles for as1, as2, as3, and as4) with sequence sv30 shown in the insets.
}
   \label{fig12}
\end{figure}

\vfill\eject

\begin{figure}
   \centering
   \includegraphics[width=\columnwidth]{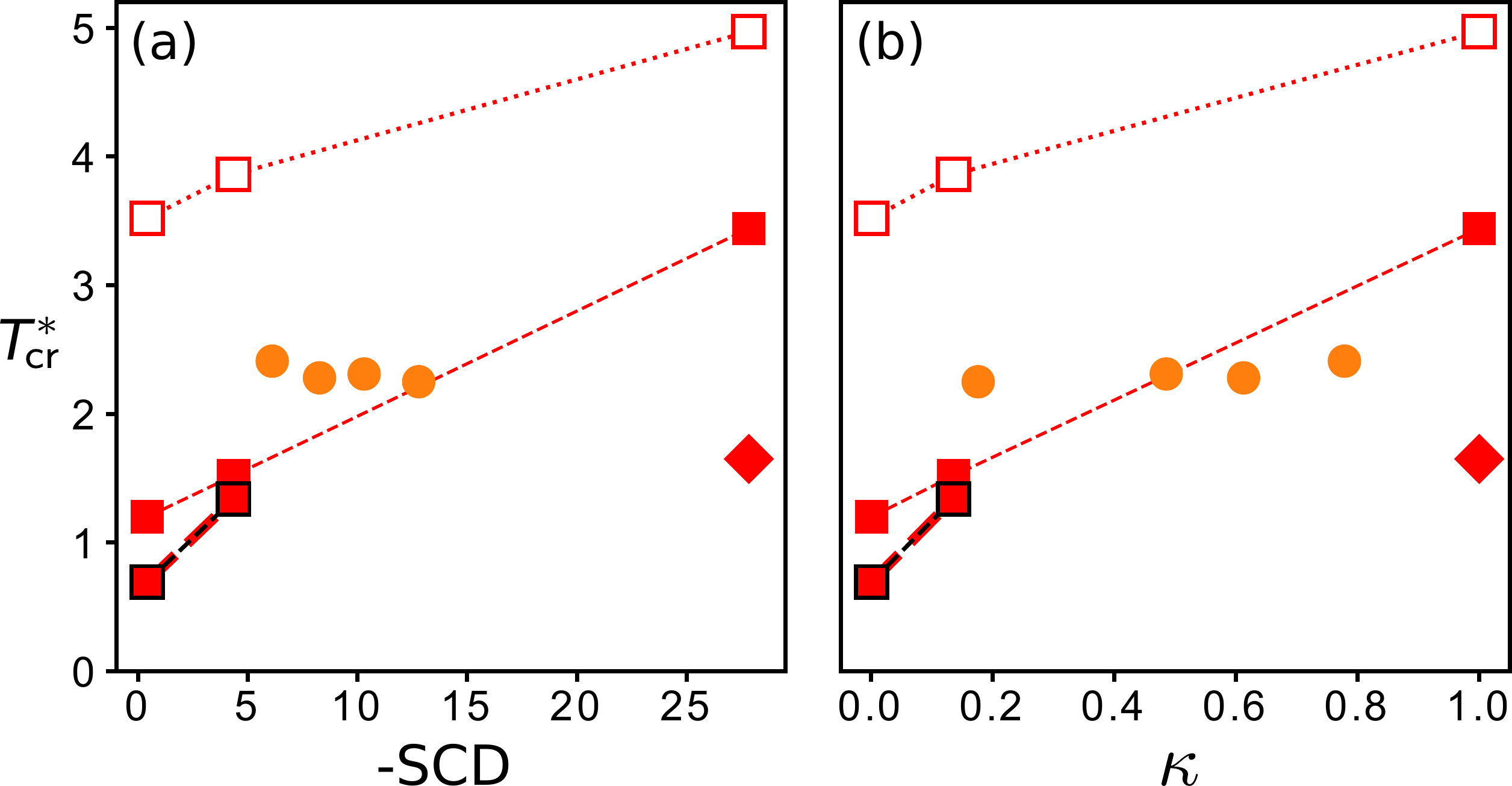}
   \caption{
Dependence of explicit-chain-simulated LLPS propensity on sequence charge
pattern parameters.
The critical temperatures, $T^*_{\rm cr}$'s, of 50-residue polyampholytes
simulated in the present explicit-chain continuum model are plotted against
$-$SCD (a) and $\kappa$ (b) for the ``with LJ'' potential in Fig.~4a 
(open squares for sv1, sv15, and sv30), the ``with 1/3 LJ'' potential in 
Fig.~4b (filled squares for the three ``sv'' sequences, filled circles
for as1, as2, as3, and as4), and the ``with hard-core repulsion'' potential 
in Fig.~4c (diamonds for sv30). The $T^*_{\rm cr}$'s of sv1 and sv15
from our previous explicit-chain lattice model simulation\cite{suman1}
are also plotted for comparison (red-filled black squares).
Dashed and dotted lines joining data points simulated using the same
models for sv1, sv15, and sv30 as well as for sv1 and sv15 are merely 
guides for the eye.
}
   \label{fig13}
\end{figure}

\vfill\eject

\begin{figure}
   \centering
   \includegraphics[width=\columnwidth]{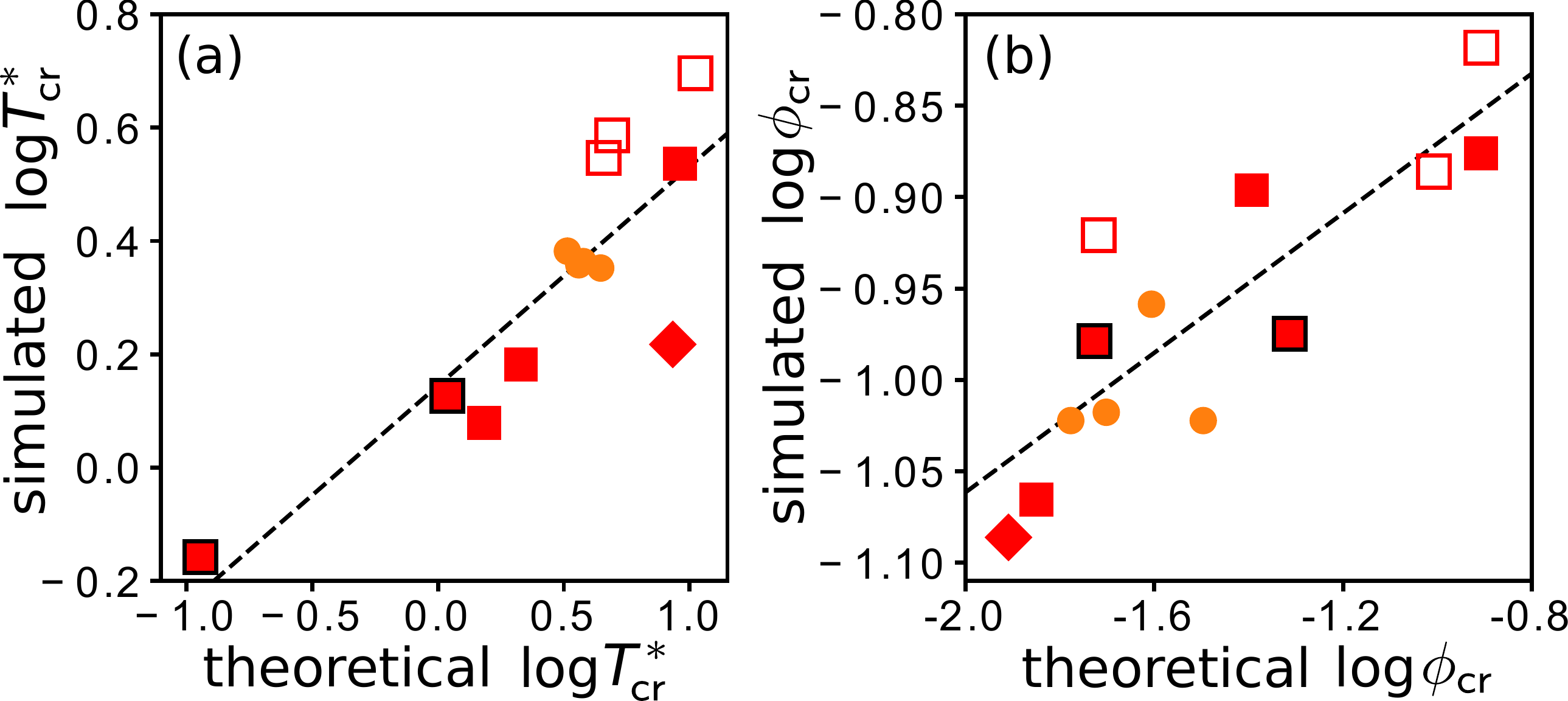}
   \caption{
Comparing RPA-predicted and explicit-chain-simulated phase properties of
polyampholytes.
Simulated logarithmic ($\log = \log_{10}$) critical temperature
$T^*_{\rm cr}$ (a) and critical volume fraction $\phi_{\rm cr}$ (b)
are compared against their theoretical counterparts predicted using
RPA or RPA+FH theories. In this figure,
$\phi_{\rm cr}$'s for the results simulated in the present study
are identified with the simulated $\rho_{\rm cr}$'s 
(see text and Table~2 for details). The same symbols as those in Fig.~13
are used to specify the sequences and simulation models for the
data points. The dashed least-squares regression 
lines are fitted using all the plotted data points and represent approximate 
power-law correlations of $T^*_{\rm cr}$ or $\phi_{\rm cr}$ between 
theory and simulation. The lines are given by
$\log(T^*_{\rm cr, sim}) = 0.387\log(T^*_{\rm cr, thr}) + 0.145$ in (a)
and $\log(\phi_{\rm cr, sim}) = 0.191\log(\phi_{\rm cr, thr}) - 0.680$ in (b)
where simulated and theoretical quantities are indicated, respectively, by the 
subscripts ``sim'' and ``thr''. The Pearson correlation coefficients
are $0.843$ for (a) and $0.846$ for (b). 
}
   \label{fig14}
\end{figure}

\vfill\eject

\begin{figure}[t]
\centering
   \includegraphics[width=\columnwidth]{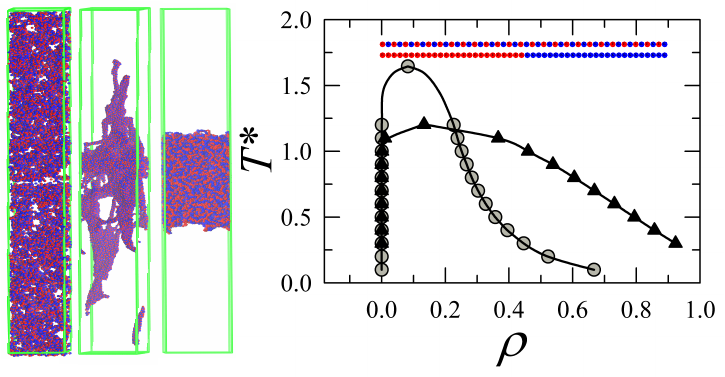}
{\large\bf Graphical Abstract}
\end{figure}

\vfill\eject

 \vfill\eject
\end{document}